\documentclass[trackchanges]{aastex701}

\usepackage{graphicx}
\usepackage{caption}
\usepackage{subcaption}
\usepackage{float}
\usepackage{morefloats}

\extrafloats{200}
\begin{document}

\title{Chromospheric and Transition Region Responses of activities at the base of Coronal Plumes}

\author[0009-0002-8422-4310]{Biswanath Malaker}
\affiliation{Inter University Centre for Astronomy and Astrophysics, Pune, India - 411007}
\email{biswanath@iucaa.in}

\author[0000-0002-9253-6093]{Vishal Upendran}
\affiliation{SETI Institute, Mountain View, CA, USA - 94043}
\affiliation{Lockheed Martin Solar and Astrophysics Laboratory, Palo Alto, CA, USA - 94304}
\email{vupendran@seti.org}

\author[0000-0003-1689-6254]{Durgesh Tripathi}
\affiliation{Inter University Centre for Astronomy and Astrophysics, Pune, India - 411007}
\email{durgesh@iucaa.in}

\begin{abstract}

We present the identification of the co-spatial and co-temporal small scale jets -- sometimes also named as jetlets -- at the base of the coronal hole plume associated with chromospheric and transition region (TR) flows. We identified 19 jets with coordinated observations recorded on March 19, 2016 using the Solar Dynamics Observatory (SDO) and the Interface Region Imaging Spectrograph (IRIS). Our analysis of \ion{Si}{4} 1393.7~{\AA} line suggests blueshifts mostly in the range of 10 to 33~km~s$^{-1}$, but few less than -10~km~s$^{-1}$, while analysis of co-spatial \ion{Mg}{2}~h~\&~k lines show both blueshifts and redshifts in the range of {--}9 to 17~km~s$^{-1}$. We observed highly asymmetric, enhanced intensity in \ion{Si}{4} line during these events. The upflows observed in transition region through \ion{Si}{4} is strongly correlated with chromospheric downflows observed in \ion{Mg}{2}. We interpret these results as a signature of interchange reconnections creating bi-directional flows -- giving rise to upflows (downflows) in transition region (chromosphere).

\end{abstract}

\section{Introduction} \label{intro}
Plumes are long-lasting, elongated structures mainly observed inside coronal holes due to their enhanced density but cooler background plasma~\citep{ahmad-1977,wilhelm-1998,wilhelm-1998a,delzanna-2003,gabriel-2003}. They were initially observed in white light \citep{vandehulst-1950,saito-1958,saito-1965} and later in EUV \citep{ahmad-1977} and X-rays \citep{ahmad-1978} inside polar coronal holes, extending up to several solar radii~\citep{newkirk-1968,deforest-2001}. They have lifetimes ranging from hours to weeks, typically lasting $\sim$20 hours~\citep{lamy-1997,withbroe-1991}.

Plumes are rooted in strong, unipolar magnetic flux regions, often coinciding with supergranular boundaries~\citep{harvey-1965,newkirk-1968,fisher-1995,deforest-2001a,deforest-1997, avallone_plumes} where the magnetic field extends like a funnel into the upper atmosphere. Their footpoints often contain coronal bright points~\citep[CBPs;][]{delzanna-2003,pucci-2014,cho-2021, weitz-2025}. There are other classes of plumes that do not contain CBPs but are often associated with jets and jetlets~\citep{raouafi-2008,malaker-2024}.

Plumes are believed to be sustained, over their lifetimes, due to small-scale jets and possibly other transients originating at the footpoints~\citep[see e.g.,][]{uritsky-2021,malaker-2024,raouafi-2008,raouafi-2014}.~\cite{malaker-2024} found a strong correlation between the brightness of the plume and the underlying emerging magnetic field and suggested that a higher number of reconnection events are possible when the flux is higher, which feeds more mass and energy into the plume, increasing the brightness. Such a correlation was also observed by~\citep{uritsky-2021} between the number of internal structures and average plume brightness; these features possibly act as conduits to the solar wind~\citep{dwivedi-2015,walker-1993,gabriel-2003,velli-2011,wang-1994,delzanna-1998}.

Although significant studies have been done to understand plumes inside the corona at a temperature of $10^6$~K over the few decades, few indirect studies have been conducted to understand the chromospheric/TR counterparts of the coronal plumes in the last decades \citep{wang-1995,allen-1997,hassler-1997,delzanna-2003}. One of the first of detailed studies came from~\cite{tian-2014a}, who showed the ubiquitous presence of chromospheric/transition region (TR) network jets inside the coronal hole with speeds from 80~km/s to 250~km/s, with ~\cite{samanta-2015,jiao-2015} reporting propagating disturbances inside plumes, co-temporal and co-spatial with chromospheric/TR spicules. \citet{bose-2025} observed spicular activities at the base of the coronal hole plume. \citet{panesar-2018}, through imaging analysis showed formation of jetlets at the base of the plumes through magnetic reconnections. \cite{pant-2014} compared the \ion{Si}{4} light curve and propagating disturbances of SJI 1330 (both representative of TR) using observations of the Interface Region Imaging Spectrograph \citep[IRIS;][]{iris-depontieu}, with propagating disturbances in coronal channels. The authors found a good correlation between transition region outflows and coronal outflows. They suggested that transient brightenings and jetlets in the upper chromosphere and lower transition region (TR) may be observed as propagating disturbances inside plumes.~\cite{cho-2023} observed a correlation of periods between type II spicules seen in the chromospheric ($H_\alpha$) channel and propagating disturbances seen inside the plume (AIA 171), which suggests that the origin of the outflow inside the base of the coronal plume is in the chromosphere.

In this paper, we aim to study the dynamics of the lower atmospheric counterparts of a coronal plume. The rest of the paper is structured as follows. In \S~\ref{data}, we describe the observations and data analysis tools that are used in this paper. We discuss the data analysis and results in \S~\ref{res}. Finally, we summarize and discuss our results in \S~\ref{discussions}.
\section{Observations and Data}~\label{data}
\subsection{Observations}~\label{observations}
Our primary aim is to study the lower solar atmospheric response of coronal plumes. For this purpose, we identify a plume in a coronal hole near the disk center, observed on March 19, 2016. The data used in this study were recorded by the Atmospheric Imaging Assembly~\citep[AIA;][]{aia-lemen} and the Helioseismic and Magnetic Imager \citep[HMI;][]{hmi-schou}, onboard the Solar Dynamics Observatory \citep[SDO;][]{sdo-pesnell}, and Interface Region Imaging Spectrograph (IRIS) \citep{iris-depontieu}.

The AIA observations are used to study the time evolution of the plume footpoint in the coronal channels. AIA records full-disk images of the Sun in seven EUV passbands centered at 94~{\AA}, 131~{\AA}, 171~{\AA}, 193~{\AA}, 211~{\AA}, 304~{\AA}, and 335~{\AA}, with a time cadence of $\approx$12~s and a pixel size of $\approx$0.6{\arcsec}, covering a broad range of temperatures \citep{o'dwyer-2010}. It also takes images in 1600~{\AA}, 1700~{\AA}, and 4500~{\AA}, albeit at a lower cadence. The line-of-sight (LOS) magnetograms from HMI are used to study the structure and evolution of the photospheric magnetic flux density. HMI provides full-disk LOS magnetograms at a temporal resolution of $\approx$45~s with a pixel size of $\approx$0.5{\arcsec}.

The IRIS observations are used to study the spectral characteristics in the chromosphere and transition region (TR). IRIS provides Slit-jaw Images (SJI) centered around \ion{C}{2}~1335~{\AA}, \ion{Si}{4}~1400~{\AA}, \ion{Mg}{2}~k~2796~{\AA}, and 2832~{\AA} wavelengths. These images can have a cadence as high as 5~s. The spectral windows covered by the IRIS spectrograph are 1332--1358~{\AA}, 1389--1407~{\AA}, and 2783--2834~{\AA}. These spectra are taken with a slit width of 0.33{\arcsec} and a height of 175\arcsec. These spectral scans can have a cadence as high as 3~s. Together with SJI  and spectra, IRIS covers a temperature range of 5000~K to 10~MK.

For this investigation, we selected three spectral lines: \ion{Si}{4}~1393~{\AA} ($\log T/K \approx 4.8$), which is sensitive to TR plasma, and \ion{Mg}{2}~k~2796~{\AA} \& \ion{Mg}{2}~h~2803~{\AA} ($\log T/K \approx 4$), which probe the upper chromosphere~\citep{iris-depontieu}. Within the IRIS observational program, we acquired slit-jaw images in the 1400~{\AA} channel. The plume region was observed by IRIS on March 19, 2016, between 12:00--18:00~UT (obsid: 3623010639) with a FOV of 15\arcsec$\times$119\arcsec and an exposure time of 14.99~s and step cadence of 16.5 s with raster cadence of 264s. For this analysis, we utilized three datasets spanning 12:14--14:35~UT, 14:57--16:12~UT, and 16:39--17:32~UT. The SJIs in 1400~{\AA} were recorded at a 22~s cadence, while those in 2832~{\AA} were recorded at a 66~s cadence, with a FOV of 119\arcsec$\times$119\arcsec. We applied standard processing techniques to the data by co-aligning all AIA, HMI, and IRIS Level-2 data. AIA and HMI data were corrected for roll angle. AIA data were rescaled to 0.6 arcsec pixel scale. SJI 1400 images were corrected for roll angle and also aligned with AIA 1600 images using cross correlation method without modifying SJI pixel scale. SJI has pixel scale of 0.166\arcsec along solar x and y directions. As the IRIS slit-jaw imager and spectrograph share a common slit and pointing solution, this alignment applies equally to the \ion{Si}{4} and \ion{Mg}{2} raster spectra.

\begin{figure}[H]
    \centering
    \includegraphics[width=0.9\textwidth]{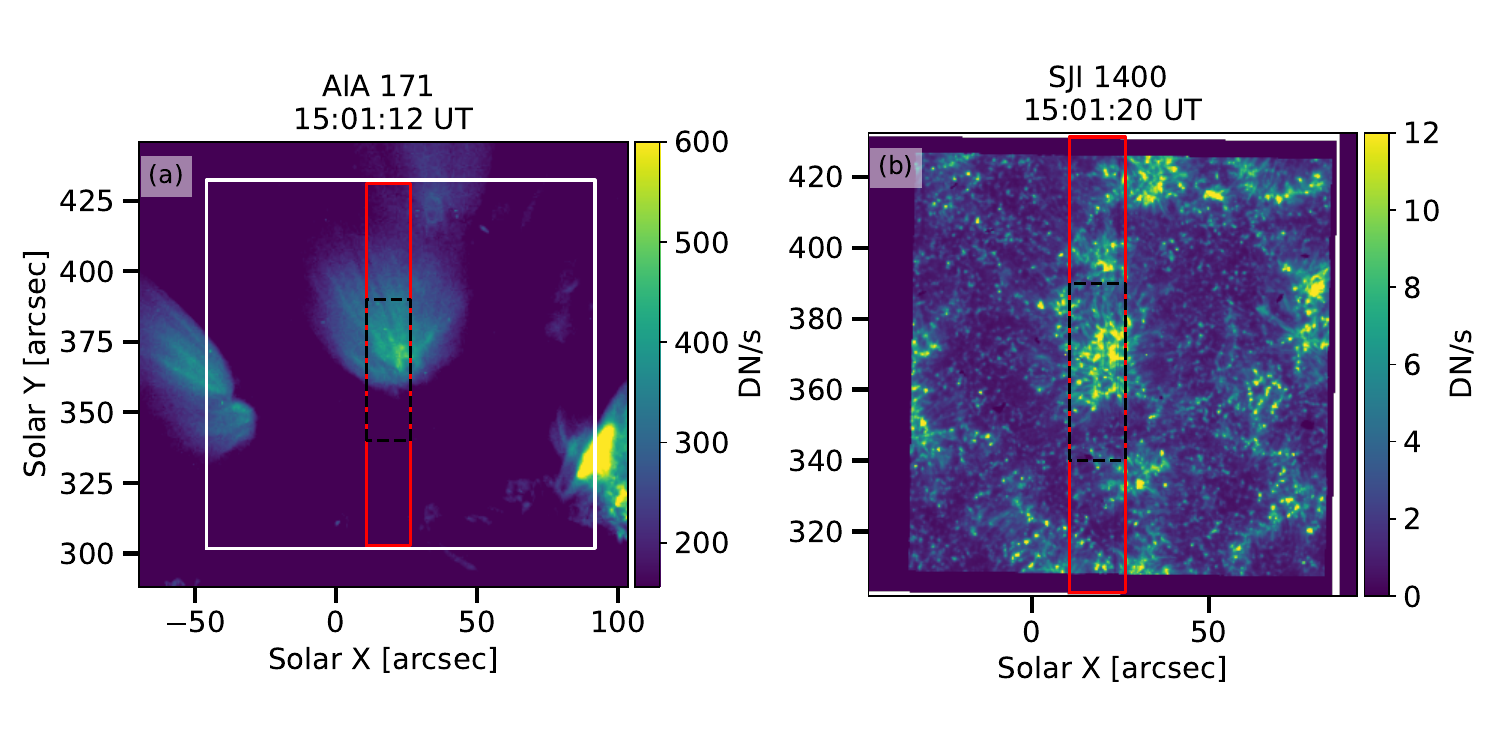}
    \caption{Panel (a): A coronal plume observed in AIA 171~{\AA}. The over-plotted white box shows the IRIS SJI field of view (FOV). Panel (b): IRIS SJI taken in 1400~{\AA} channel. The over-plotted red box in both panels denotes the IRIS raster FOV. The dashed black box locates the part of the raster FOV that is used in this work.}
    \label{fig:context}
\end{figure}

Figure~\ref{fig:context} displays the plume that we study in this paper. In panel (a), we show the plume observed in 171~{\AA} recorded by AIA, while in panel (b), we show the IRIS SJI 1400 image at the nearest timestamp. The over-plotted white box in panel (a) locates the FOV of the SJI. The regions enclosed by the red box in both panels show the IRIS raster FOV. The region shown by the black dashed box---primarily covering the footpoint of the plume---is considered for further studies.

\subsection{Spectral Fitting and Feature Generation}\label{fitting_strategy}

To enhance the signal-to-noise ratio (SNR), we perform a 2-pixel averaging of the spectra along the y-direction for both the \ion{Si}{4} and the \ion{Mg}{2} lines. The photon noise was estimated following standard IRIS calibration practices\footnote{\href{https://hesperia.gsfc.nasa.gov/ssw/iris/idl/nrl/iris_getwindata.pro}{$iris\_getwindata.pro$}}. We estimated the noise for the binned pixels by propagating the individual pixel uncertainties and adding them in quadrature, i.e., $\sqrt{\sum err_i^2}/N$.

From the \ion{Si}{4} line, we estimate the intensity, velocity, and line width of one or multiple Gaussian components. We estimate these parameters using the \texttt{curve\_fit} method provided by \texttt{SciPy} \citep{scipy} for fitting \ion{Si}{4} spectra, which utilizes the Trust Region Reflective (TRR) algorithm \citep[see][for details of the algorithm]{branch-1999}, considering a rest wavelength of 1393.755~{\AA}. The boundary conditions for the amplitude are defined as [0, max (Intensity)] of the line The bounds for mean and standard deviation are kept as [1393.6~\AA{}, 1393.9~\AA{}]. We consider a linear background, with a slope bounded by [-1, 1], and a constant [0,1]. We take a number of initial guesses with varying mean and standard deviation (0.01 \AA{} to 0.3 \AA{} uniformly) and selected the best fit considering the lowest $\chi^2_{red}$. To perform the double Gaussian fits, we take a range of initial parameters surrounding the best parameters obtained using single Gaussian fits. The amplitude is varied from 20\% to 80\% of single Gaussian fit, while the mean varied around $\pm25$\% from single Gaussian fit, and the standard deviations of {$\pm$}30\%. We have taken initial slope and constant for linear background fit as 0.1 and 0.01 for both single and double gaussian fitting. Finally we consider the best fit by considering the lowest $\chi^2_{red}$ of the double Gaussian fitting. The bounds for double Gaussian are the same for single Gaussian fitting.

The \ion{Mg}{2} lines are known to form under optically thick conditions \citep{leenaarts-2013}. Hence, Gaussian fits by themselves do not directly correspond to plasma properties in the solar atmosphere. However, \citet{leenaarts-2013,leenaarts-2013a,pereira_2013} demonstrated that specific features extracted from the \ion{Mg}{2} lines are correlated with the properties of the solar chromosphere. The features of interest considered here are: k2v/h2v (blue-ward emission peak), k2r/h2r (red-ward emission peak), and k3/h3 (central absorption dip). The Doppler shift of k3 and h3 corresponds to the plasma velocity in the upper chromosphere, while the average Doppler shift of the peaks corresponds to the plasma velocity in the middle chromosphere.

To extract features from the \ion{Mg}{2} h and k spectra, we first perform a cubic spline interpolation of the spectra onto a high-resolution wavelength grid. To suppress noise without distorting spectral features, the spectra are averaged with a Gaussian kernel with a width of 1 pixel (2 pixels were used for two specific jets in Table~\ref{tab:jet_si_mgii} to ensure features were correctly identified). We then use the \texttt{find\_peaks} method from the \texttt{scipy.signal} library to detect maxima and minima, through which we define the blue and red peaks and the central reversal. We note that our spectra of interest typically have two peaks and a central reversal; therefore, we do not consider any profiles with more than two peaks. To estimate the line shifts, we use rest wavelengths of 2803.5297192~{\AA} and 2796.3509493~{\AA} for the \ion{Mg}{2} h and k lines, respectively. We determine the line-integrated flux for the \ion{Mg}{2} h and k lines using the trapezoidal integration method from \texttt{scipy.integrate} \citep{scipy}.

This exercise of fitting the lines allows us to identify jets occurring at the base of the plume shown in Fig.~\ref{fig:context}. In this study, we consider the evolution of the plume for approximately 5 hours in AIA 171~{\AA} and the corresponding IRIS spectra, spanning 12:30~UT to 17:32~UT for the three datasets mentioned in \S~\ref{observations}.

\section{Data Analysis and Results}\label{res}
Since we intend to study jetting activity at the plume base, we utilize a multi-step definition for a \emph{jet}. We first identify pixels which show intensity enhancements in AIA 171~{\AA} as markers of jetting activity. Then, we consider all the IRIS rasters and select pixels based on the fitting of the \ion{Si}{4} line to pick out pixels corresponding to jets. We present this scheme below.

\subsection{Time Evolution of Plumes and Jets}\label{time_evolution}

\begin{figure}[!h]
\centering
\includegraphics[width=\textwidth]{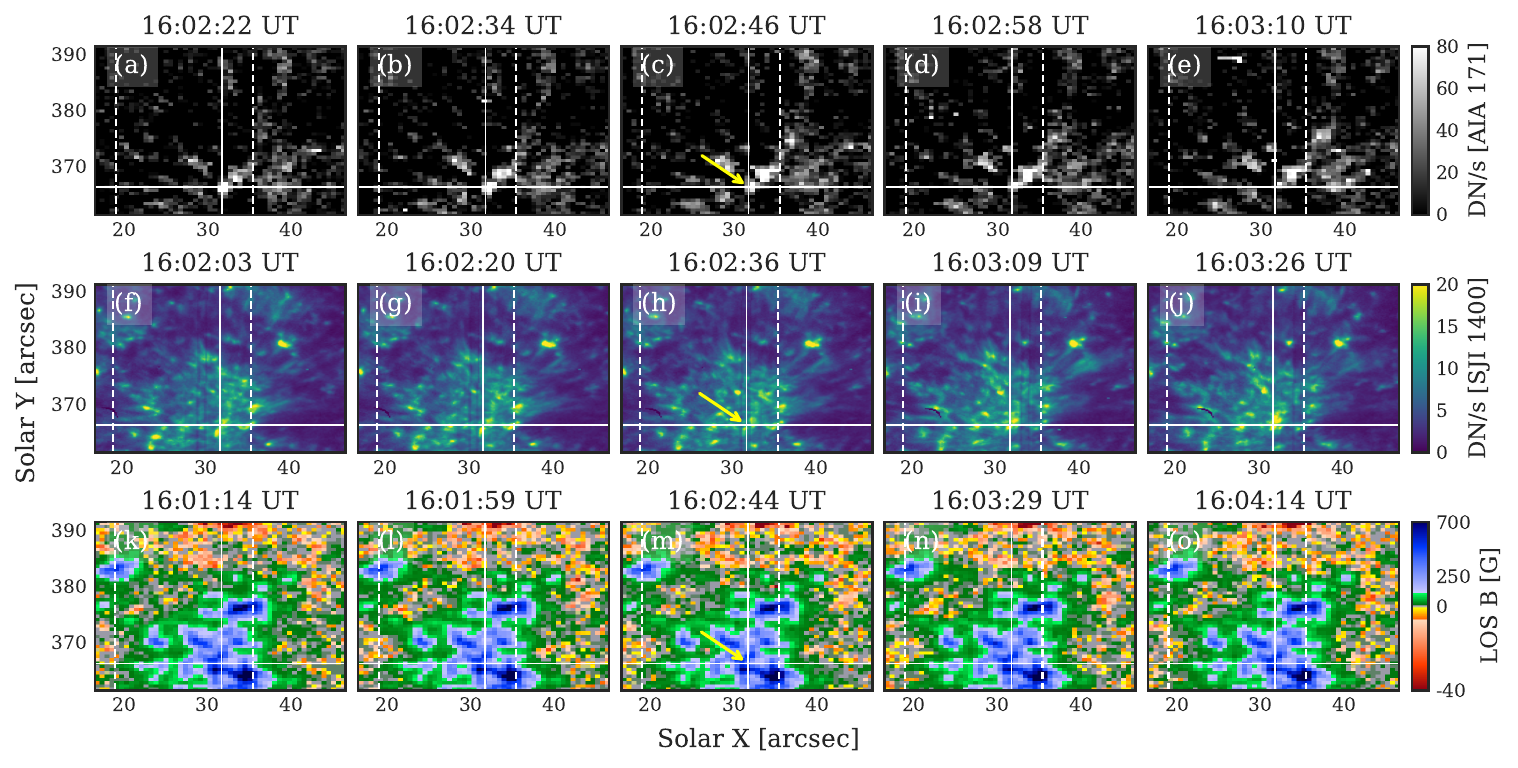}
\caption{Time evolution of the footpoint of the plume as observed by AIA 171~{\AA}. Top row: base difference images (subtracted image from a frame 120s before the middle column) taken by AIA~171{\AA}. Middle row: Corresponding near-simultaneous IRIS-SJI maps at nearest time when the spectra shown in Fig.~\ref{fig:jet06_difference_image} were recorded. Bottom row: Line of Sight (LOS) magnetic flux density obtained from HMI. The two vertical white dashed lines show the IRIS raster FOV. The white cross-hair mark locates the position of the jet, which is also demarcated by yellow arrows in the middle column.}\label{fig:jet06_sequence}
\end{figure}

We first study the time-series evolution of the plume and seek to identify intensity enhancements corresponding to jetting activity at the base of the plume. In Fig.~\ref{fig:jet06_sequence}, we display the time evolution of the plume footpoint for a short window spanning an intensity enhancement. Each column corresponds to a timestamp, while the three rows correspond to observations in AIA 171~{\AA}, SJI 1400~{\AA}, and line-of-sight (LOS) magnetic flux density. The middle column (panels c, h, m) shows the maps at the time nearest to the jet formation, with the location of the jet marked with a yellow arrow.  We note that the magnetic flux density at the base of the plume is very high ($\sim$700~G), and we do not observe any opposite-polarity magnetic field at the jet footpoint. The black dashed lines  correspond to the IRIS spectrograph FOV, within which the different spectral properties are computed.

\subsection{Raster Maps}\label{raster_maps}

We next consider the features generated from the \ion{Si}{4} and \ion{Mg}{2} lines from the IRIS rasters. For the \ion{Si}{4} line, we first perform a single-Gaussian fit as explained in \S~\ref{fitting_strategy}. However, some pixels at the base of the plume are better explained by a double-Gaussian fit instead of a single-Gaussian fit when we compare the $\chi^2_{red}$ values. Hence, of the pixels that show intensity enhancements in \S~\ref{time_evolution}, we consider the pixels that have two flow components in \ion{Si}{4}. We mask out regions that show $\chi^2_{red} \leq 1$ with a single Gaussian fit of the \ion{Si}{4} line, while only the pixels with $\chi^2_{red} > 1$ are displayed with a relevant colormap, as those lines require double-Gaussian fitting.

\begin{figure}[H]
    \centering
    \includegraphics[width=0.9\textwidth]{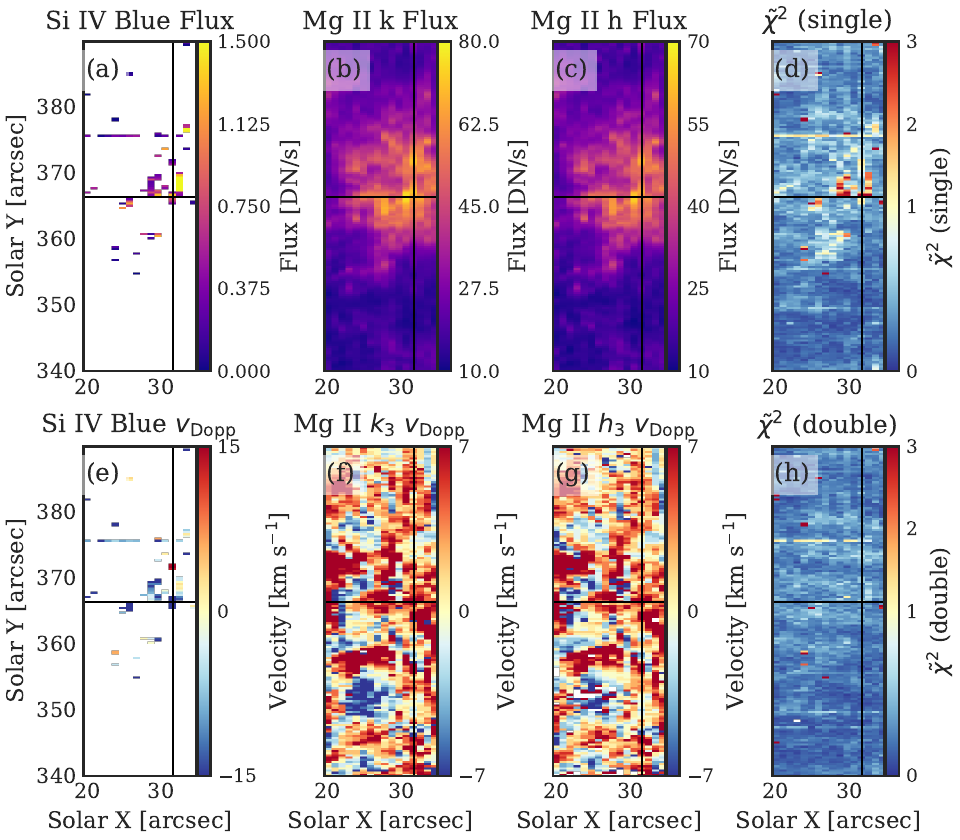}
    \caption{IRIS raster maps for various quantities after applying Gaussian fitting and Mg II feature identifications. Top row: Intensity maps obtained in \ion{Si}{4} blue component (panel a), \ion{Mg}{2}~k~\&~h (panels b \& c) and reduced chi-square map for single gaussian fit of the \ion{Si}{4} line (panel d). Bottom row: Doppler maps obtained in  \ion{Si}{4} blue component (panel e), \ion{Mg}{2}~k~\&~h (panels f \& g). Panel (h) displays the reduced chi-square map for double gaussian fitting of the \ion{Si}{4} line. The cross hair is same as that shown in Fig.~\ref{fig:jet06_sequence}. panels (a) and (e) shows the color for only those pixels for which reduced chi square for the single gaussian fitting in the \ion{Si}{4} line is greater than one.}
    \label{fig:jet06_raster}
\end{figure}

We consider one example raster to demonstrate our jet identification strategy. In Fig.~\ref{fig:jet06_raster}, we display the fluxes in the blue component of the double-Gaussian fit of \ion{Si}{4}, the \ion{Mg}{2} k and h lines, and the $\chi^2_{red}$ of the single-Gaussian fit of \ion{Si}{4} in the top row. In the bottom row, we show the Doppler shift of the blue component of the double-Gaussian fit of the \ion{Si}{4}, the Doppler shift of \ion{Mg}{2} k3 and h3, and the $\chi^2_{red}$ of the double-Gaussian fit to the \ion{Si}{4} spectra. We have masked out the regions (pixels with $\chi^2_{red} \leq 1$), which are already explained by a single-Gaussian fit, depicted by the blue color in Fig.~\ref{fig:jet06_raster}d. We find that the \ion{Si}{4} spectra are more complex at the base of the plume, while the \ion{Mg}{2} quantities do not show any drastic signatures of the plume, albeit a slight enhancement in the intensity. From the \ion{Si}{4} blue-component, we only consider pixels which show a blueshift and mark them with the cross-hair in all panels.

\subsection{Jet Definition and Identification}\label{jet_id}

\begin{figure}[!ht]
    \centering
    \includegraphics[width=0.8\textwidth]{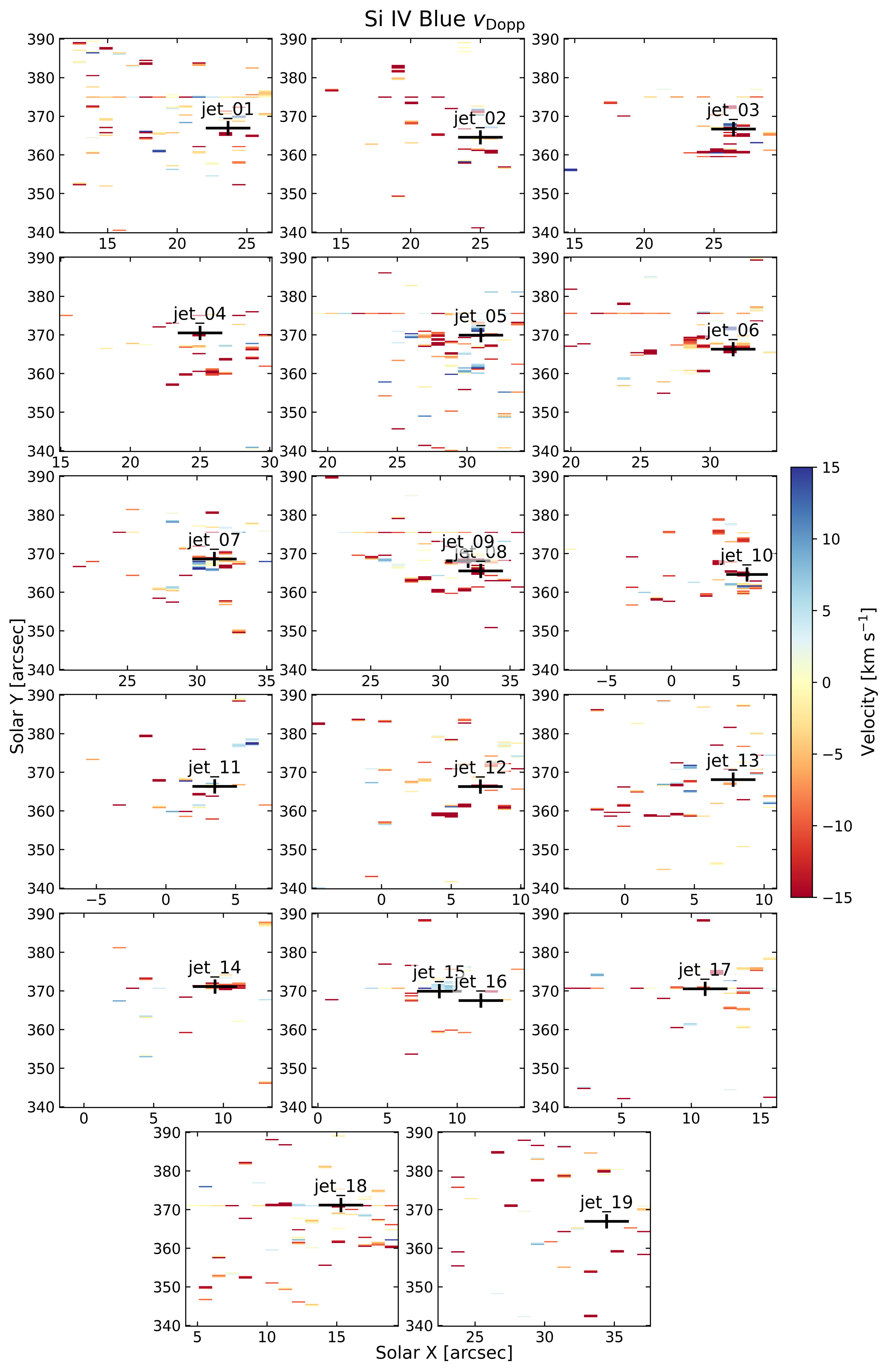}
    \caption{Jet locations for different raster scans in velocity of the blue component of \ion{Si}{4}. Jet locations are shown by horizontal bars spanning the pixels. Only pixels with $\chi^2_{\mathrm{red}} > 1$ for a single Gaussian fit are displayed.}
    \label{fig:all_raster_jets}
\end{figure}

A pixel is defined to contain a jet if we see i) collimated rapid upflow at the base of the plume in AIA 171~{\AA}, and ii) it is explained by two Gaussian components in \ion{Si}{4}, and shows a clear blue-shifted component. The second choice was primarily motivated by paper \citet{tian-2014a} who observed high speed outflows in the transition region network jets and further investigated by other authors \citep{narang-2016,kayshap-2018,gorman-2022}. We further confirm the presence of jet by manually considering difference images between the current frame and a frame from 2 minutes earlier. We perform this analysis for all of our IRIS rasters, and display them in Fig.~\ref{fig:all_raster_jets}, where we identify 19 jets in this raster sequence.  The properties of these jets in different spectral lines are listed in Table~\ref{tab:jet_si_mgii} and the locations of these jets on the rasters are provided in Figure~\ref{fig:all_raster_jets}. All jets identified here show a blueshifted \ion{Si}{4} component because this was a criterion of our jet definition, not because other configurations are physically excluded.

\begin{deluxetable}{cccccccccc}[H]
\tabletypesize{\footnotesize}
\tablewidth{0pt}
\tablecaption{ \ion{Si}{4} Gaussian fit parameters and Mg II k and h line Doppler shifts for all jets
and AIA and raster scan time offsets.
\label{tab:jet_si_mgii}}
\tablehead{
\colhead{jet id}   &   \colhead{$v^{Si}_{blue}$}   &   \colhead{$v^{Si}_{red}$}   &   \colhead{FWHM$_{blue}$}   &   \colhead{FWHM$_{red}$}   &   \colhead{$v_{k3}$}   &   \colhead{$v_{k2}$}   &   \colhead{$v_{h3}$}   &   \colhead{$v_{h2}$}   &   \colhead{Time offset}\\
\colhead{}   &   \colhead{km s$^{-1}$}   &   \colhead{km s$^{-1}$}   &   \colhead{km s$^{-1}$}   &   \colhead{km s$^{-1}$}   &   \colhead{km s$^{-1}$}   &   \colhead{km s$^{-1}$}   &   \colhead{km s$^{-1}$}   &   \colhead{km s$^{-1}$}   &   \colhead{s}
}
\startdata
jet\_01   &   -27.3   &   2.2   &   24.7   &   27.5   &   -0.8   &   0.7   &   -1.2   &   0.2   &   9.12\\
jet\_02   &   -12.8   &   19.9   &   45.0   &   27.3   &   12.2   &   2.2   &   10.6   &   2.2   &   7.2\\
jet\_03   &   -21.9   &   12.1   &   35.1   &   36.5   &   -0.6   &   4.8   &   2.2   &   3.6   &   5.31\\
jet\_04   &   -20.3   &   12.5   &   42.7   &   27.7   &   8.2   &   3.7   &   7.6   &   2.2   &   13.28\\
jet\_05   &   -16.5   &   11.5   &   26.5   &   28.8   &   3.2   &   5.7   &   0.7   &   5.1   &   10.65\\
jet\_06$^*$   &   -24.3   &   -6.9   &   21.9   &   50.5   &   9.2   &   6.7   &   8.6   &   6.1   &   9.69\\
jet\_07   &   -17.4   &   11.5   &   31.2   &   37.0   &   4.2   &   0.7   &   2.5   &   0.4   &   13.24\\
jet\_08   &   -26.7   &   -9.8   &   19.1   &   59.6   &   -0.8   &   2.2   &   -0.3   &   2.7   &   7.76\\
jet\_09   &   -30.3   &   -20.2   &   28.3   &   62.0   &   10.2   &   11.7   &   10.6   &   13.0   &   12.25\\
jet\_10   &   -16.0   &   15.4   &   28.9   &   28.6   &   1.2   &   6.2   &   2.7   &   4.7   &   9.69\\
jet\_11   &   -5.9   &   22.5   &   36.1   &   26.3   &   4.2   &   -0.3   &   1.7   &   0.7   &   5.96\\
jet\_12   &   -13.7   &   3.2   &   43.8   &   20.0   &   1.2   &   -1.8   &   -0.3   &   -1.7   &   9.7\\
jet\_13   &   -3.4   &   21.7   &   23.5   &   27.5   &   17.2   &   4.2   &   16.6   &   4.0   &   8.75\\
jet\_14   &   -32.9   &   14.0   &   33.4   &   31.7   &   12.2   &   7.7   &   10.6   &   6.6   &   9.46\\
jet\_15   &   -11.1   &   9.7   &   36.9   &   50.2   &   8.2   &   2.7   &   3.7   &   2.2   &   4.41\\
jet\_16   &   -15.3   &   -13.6   &   20.9   &   44.9   &   1.2   &   0.2   &   2.7   &   0.7   &   15.03\\
jet\_17   &   -26.4   &   -2.0   &   8.1   &   55.3   &   13.2   &   4.7   &   12.5   &   3.7   &   10.07\\
jet\_18$^*$   &   -17.8   &   -7.5   &   41.8   &   16.9   &   -6.1   &   -4.7   &   -8.7   &   -4.6   &   12.84\\
jet\_19   &   -14.4   &   -7.7   &   47.6   &   28.9   &   -7.7   &   -4.7   &   -8.1   &   -4.2   &   10.1\\
\enddata
\tablecomments{Time offset is the AIA identification-frame time minus the IRIS spectral exposure start time at the jet location; max offset $\sim$15~s.}
\end{deluxetable}

\subsection{Analysis of Two Example Jets} \label{examples}

From our listed 19 jets, we first present a detailed analysis of two jets marked by an asterisk in Table~\ref{tab:jet_si_mgii} to understand the spectral signatures in different layers of the solar atmosphere.

\subsubsection{Jet \#06: Upflow in \ion{Si}{4} but downflows in \ion{Mg}{2}:}

\begin{figure}[!ht]
    \centering
    \begin{interactive}{animation}{jet_06_AAS.mp4}
    \includegraphics[width=0.9\textwidth]{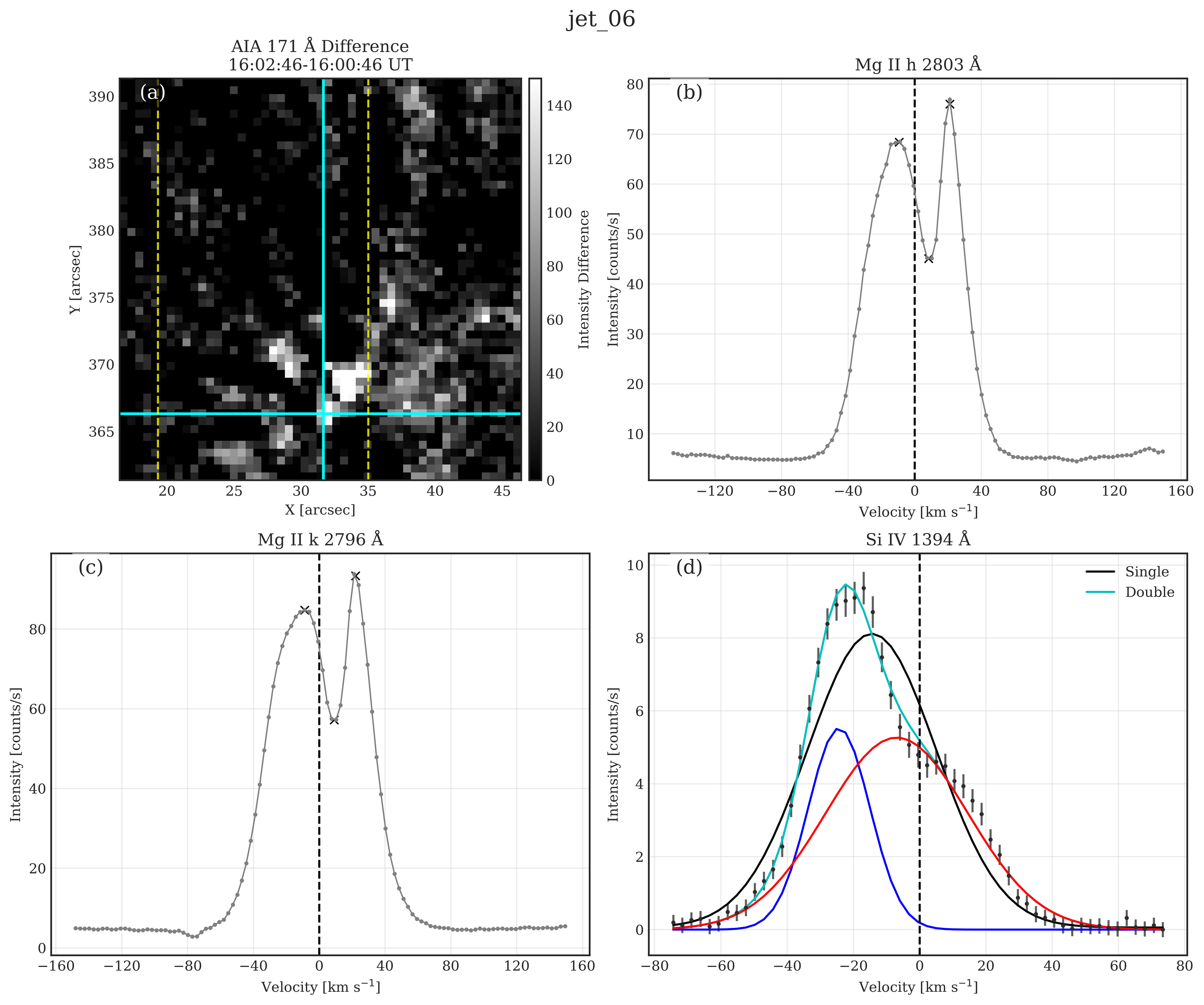}
    \end{interactive}

    \caption{Panel a: AIA 171 \AA{} base difference map at the time when the spectral line was recorded. The AIA image is subtracted from that taken 120~s earlier. Crosshair locates the position of the jet\_06 at which the spectrum was recorded and yellow dashed lines represent raster scan regions. Panel (b) and (c) show \ion{Mg}{2}~h and k lines with 'x' marks denoting peaks and dips. Panel (d) shows \ion{Si}{4} spectra over-plotted with single (black solid) and double (cyan solid) Gaussian fits. Blue and red solid lines show the blue and red components of the double gaussian fit. An animated version of this figure is available in the HTML version. The animation (4.2 s duration) consists of 20 original observation frames and shows the temporal evolution of the AIA 171 \AA\ intensity and the base difference maps. The animation focuses on the plume region to highlight the intensity enhancement at the jet location over time.} \label{fig:jet06_difference_image}
\end{figure}

Jet \#06 is an example where we find blueshifts in the blue component of \ion{Si}{4}, while we find redshifts in the \ion{Mg}{2}~k3 line profile. 
We display the identification and spectral lines of Jet \#06 in Fig.~\ref{fig:jet06_difference_image}. In panel (a), we display base difference image of the plume from two minutes prior as observed in AIA~171~{\AA}, while the cross-hair in cyan denotes the jet location. In panels (b) and (c), we show the \ion{Mg}{2} h and k spectra, respectively, where the k2v, k2r, and k3 features are located with crosses. In panel (d) we display the \ion{Si}{4} spectral line (black dots with errors), the single-Gaussian fit (black solid curve), the two Gaussian components (blue and red), and the sum of the two-component profiles (cyan). The red and blue components are defined based on their relative position, and not on the absolute Doppler shift from the rest wavelength.

From the difference images, we see a clear enhancement in the AIA 171~{\AA} intensity at the jet location inside the plume (see also the animation associated with fig~\ref{fig:jet06_difference_image}). We also find that the \ion{Si}{4} line in Fig.~\ref{fig:jet06_difference_image}d reveals a highly skewed line profile, which is best explained by a two-component Gaussian fit. The blue component of \ion{Si}{4} is at a blueshift of 24.3~km~s$^{-1}$, with a full width at half maximum (FWHM) of 21.9~km~s$^{-1}$. The red component is also blueshifted at 6.9~km~s$^{-1}$, with a FWHM of 50.5~km~s$^{-1}$. Co-spatially, we find the \ion{Mg}{2}~k3 shift to be 9.2~km~s$^{-1}$, while the \ion{Mg}{2}~h3 shift is 8.6~km~s$^{-1}$. Finally, the average peak shift of \ion{Mg}{2}~k2 is $\sim$6.7~km~s$^{-1}$, and that of \ion{Mg}{2}~h2 is 6.1~km~s$^{-1}$.

\subsubsection{Jet \#18: Upflows in \ion{Si}{4} as well as \ion{Mg}{2}:}

\begin{figure}[H]
    \centering
    \begin{interactive}{animation}{jet_18_AAS.mp4}
    \includegraphics[width=0.9\textwidth]{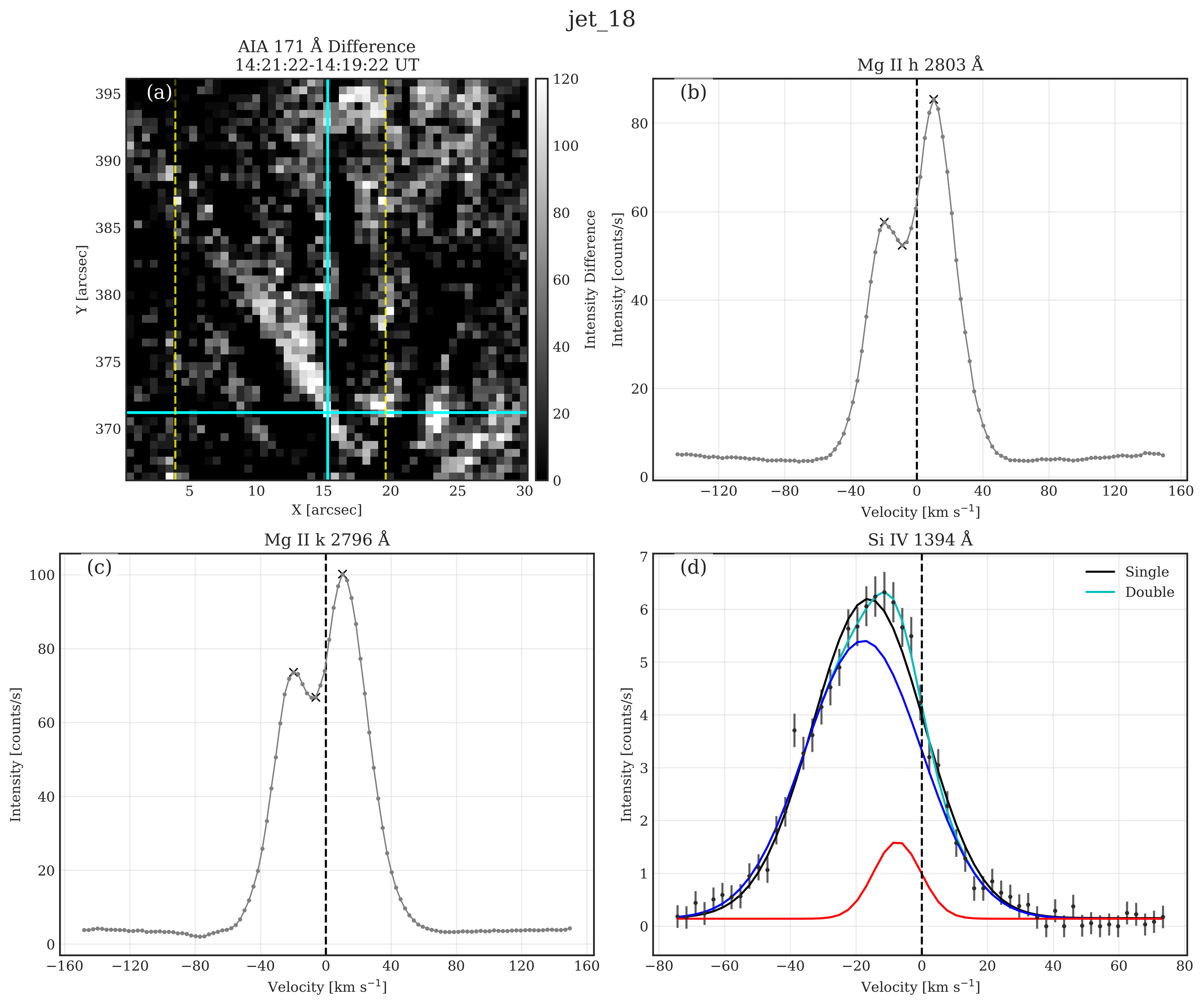}
    \end{interactive}

    \caption{Same as Fig.~\ref{fig:jet06_difference_image} but for jet18. An animated version of this figure is available in the HTML version. The animation (4.2 s duration, 20 frames) shows the temporal evolution of the AIA 171 \AA\ intensity and the base difference maps for the jet 18 at the plume region.} \label{fig:jet18_difference_image}
\end{figure}

Jet \#18 is a typical case where we find upflows in the \ion{Si}{4} blue component as well as in the \ion{Mg}{2}~h and \ion{Mg}{2}~k lines. There are only five such jets in our sample where both \ion{Si}{4} and \ion{Mg}{2}~k3 exhibit blueshifts (see Table~\ref{tab:jet_si_mgii}). 
Figure~\ref{fig:jet18_difference_image} is similar to Fig.~\ref{fig:jet06_difference_image}. Panel (a) shows the AIA~171~{\AA} difference image from two minutes prior, while the cyan crosshairs indicate the location of the jet. Panels (b) and (c) correspond to the \ion{Mg}{2} spectra, with features marked by crosses. Panel (d) corresponds to the \ion{Si}{4} spectrum (black dots with errors), the single-Gaussian fit (black solid line), the red and blue components of the double-Gaussian fit, and the net double-Gaussian fit (cyan). An animation associated with fig~\ref{fig:jet18_difference_image} is also provided that shows evolution of jet 18.
For this particular jet, the \ion{Si}{4} line has a blueshift of $17.8~\mathrm{km~s}^{-1}$ in the blue component and a blue shift of $7.5~\mathrm{km~s}^{-1}$ in the red component, with a FWHM of $41.8~\mathrm{km~s}^{-1}$ and $16.9~\mathrm{km~s}^{-1}$, respectively. The \ion{Mg}{2}~k3 and h3 lines have blue shifts of $6.1~\mathrm{km~s}^{-1}$ and $8.7~\mathrm{km~s}^{-1}$, while the average peak blue shifts of the \ion{Mg}{2}~k2 and h2 lines are $4.7~\mathrm{km~s}^{-1}$ and $4.6~\mathrm{km~s}^{-1}$, respectively.

\subsection{Correlation of Velocities}\label{correlation_section}

\begin{figure}[!ht]
    \centering
    \begin{tabular}{c}
        \includegraphics[width=0.95\textwidth]{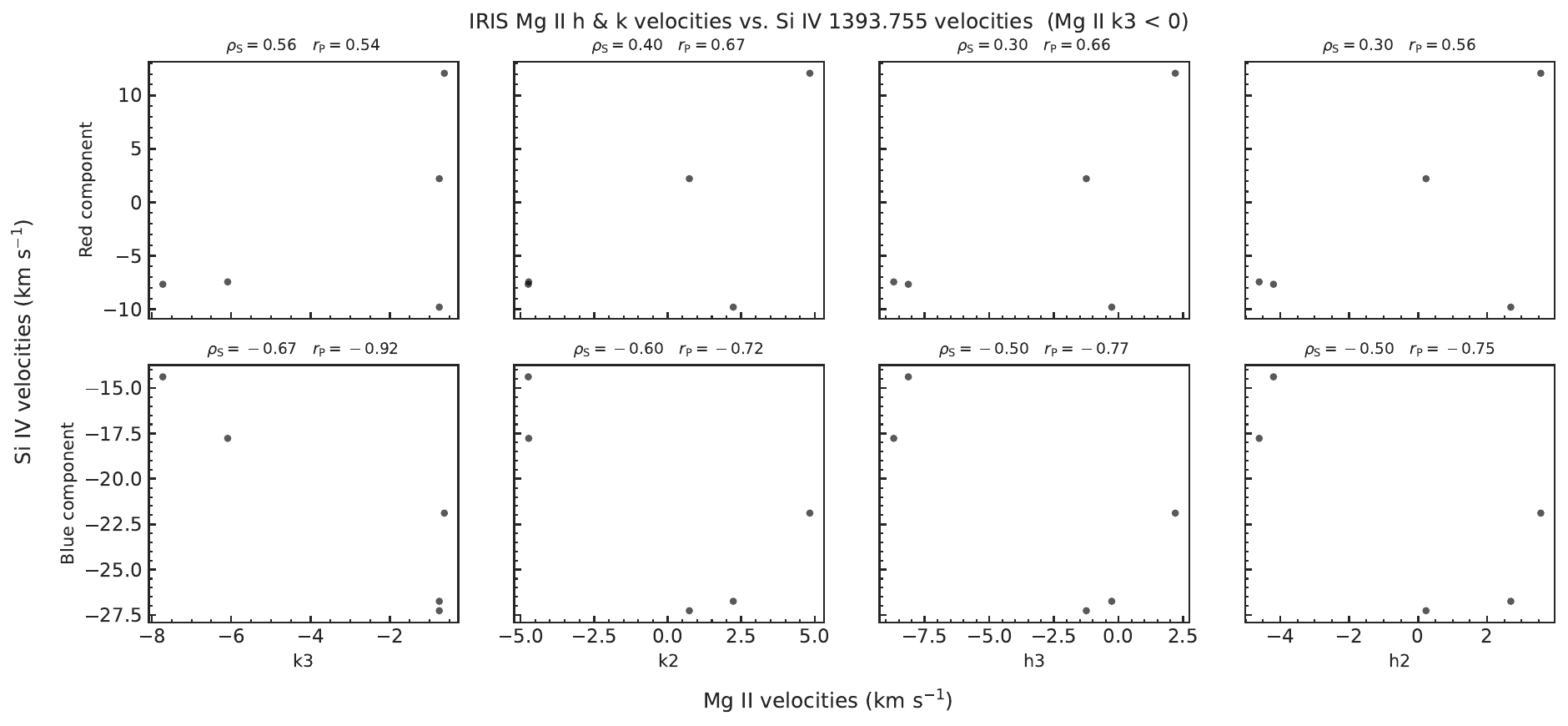} \\
        (a) \\[6pt]
        \includegraphics[width=0.95\textwidth]{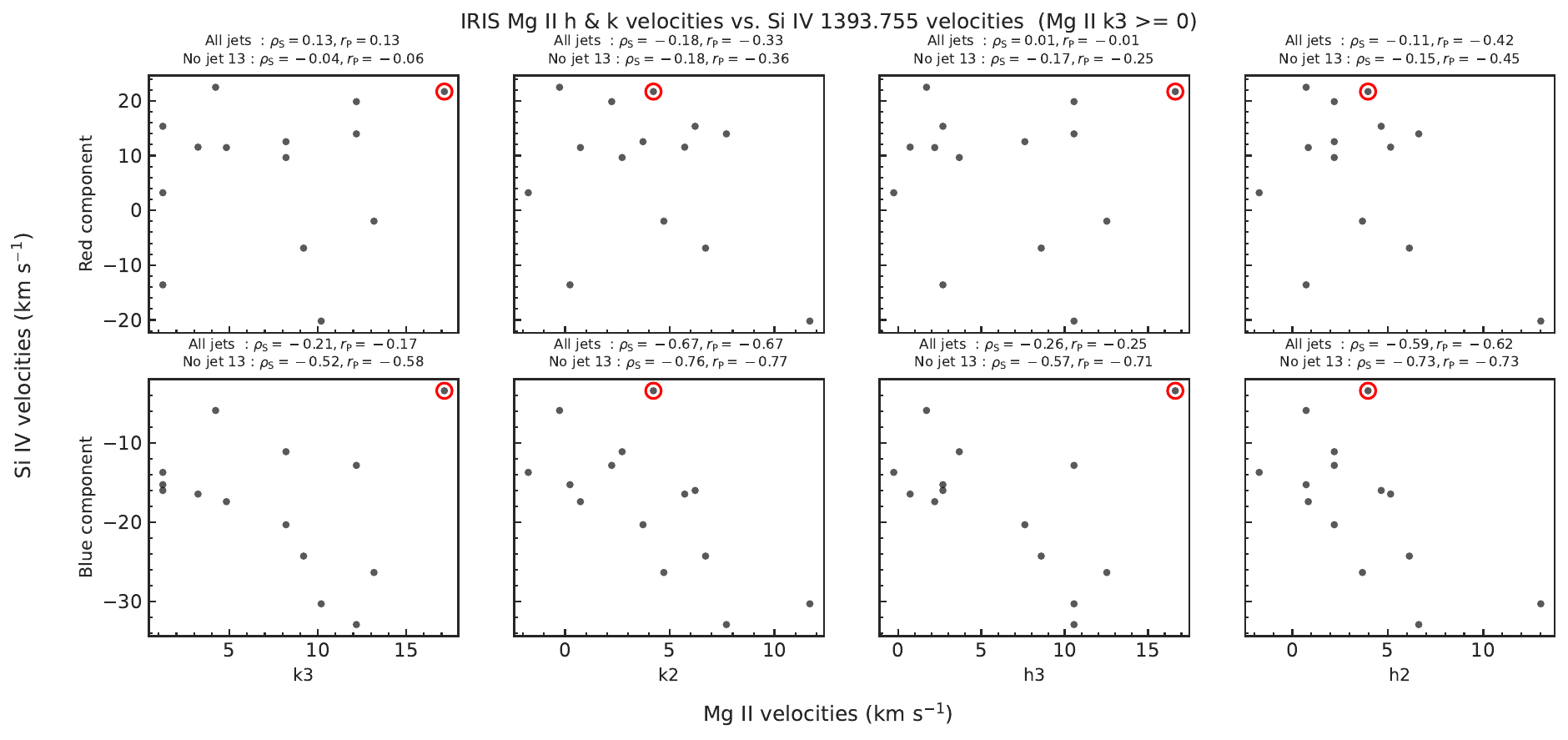} \\
        (b)
    \end{tabular}
    \caption{Correlation plot between \ion{Mg}{2}~k and h velocities and corresponding velocities in \ion{Si}{4}. (a) Blueshift in \ion{Mg}{2} k3, (b) Redshift in \ion{Mg}{2} k3. Spearman correlation coefficient ($\rho_S$) and ($r_P$) denote the Spearman and Pearson correlation coefficients, respectively. Jet 13 is marked by a red circle.}
    \label{fig:jointplot_mgk_vs_si}
\end{figure}

Having measured the properties for all the jets, as listed in Table~\ref{tab:jet_si_mgii}, we now study the relationship between the jet properties in the transition region and the chromosphere. We consider the correlation between the two components of \ion{Si}{4} line and the Doppler shift in the \ion{Mg}{2} line features in Fig.\ref{fig:jointplot_mgk_vs_si}.  We present the scatter of velocities in the red (top row) and blue (bottom) component of \ion{Si}{4} along the Y-axis in both figures. Along the X-axis, we present the Doppler shift of k3, k2, h3 and h2. In sub-figure (a), we present the correlation for blueshifts in the \ion{Mg}{2}~k3 features, while in sub-figure (b) we present the pixels showing redshift in \ion{Mg}{2}~k3. We also denote the Spearman correlation coefficient ($\rho_S$) and the Pearson correlation coefficient value ($r_p$) in the title of each panel. The Spearman correlation measures monotonicity, while the Pearson correlation measures the linearity of the relation between two variables.

From Fig.~\ref{fig:jointplot_mgk_vs_si}a, we can see that in general not many jets show blueshifts in \ion{Mg}{2}. With these small number statistics, we find that the blue component of \ion{Si}{4} is anti-correlated with \ion{Mg}{2} velocities especially for k3 and h3 lines with both spearman and pearson correlation (Fig.~\ref{fig:jointplot_mgk_vs_si}).. In contrast, the red component of \ion{Si}{4} does not show any similar correlation with the blueshifts of \ion{Mg}{2}. Once again, note that the number of data points is very small, hence, we will caution against over-interpreting these results of Fig.~\ref{fig:jointplot_mgk_vs_si}a.

The correlation plots shown in  Fig.~\ref{fig:jointplot_mgk_vs_si}(b), have better statistics. We first note that the red-component of \ion{Si}{4} does not show any strong correlation with the \ion{Mg}{2} redshifts. However, the blue component of \ion{Si}{4} shows strong anti-correlation with \ion{Mg}{2} velocities. Note that an anti-correlation in this case is for the signed quantity, i.e., the higher the velocity magnitude in \ion{Mg}{2}, the higher the velocity magnitude in \ion{Si}{4}. 

We also emphasize one event, jet-13,  which is encircled in Fig.~\ref{fig:jointplot_mgk_vs_si} b. While including that or not in the top row of Fig.~\ref{fig:jointplot_mgk_vs_si} b does not make a difference, the correlation coefficients improves significantly in the bottom row plots without jet 13. For example, for the correlation between the blue component of \ion{Si}{4} and k3 velocity, the Spearman correlation coefficient changes from -0.21 to -0.52 while the Pearson correlation coefficient changes from -0.17 to -0.58. Therefore, we study the line profile of jet 13 in a greater detail.

\begin{figure}
    \centering
    \includegraphics[width=0.8\textwidth]{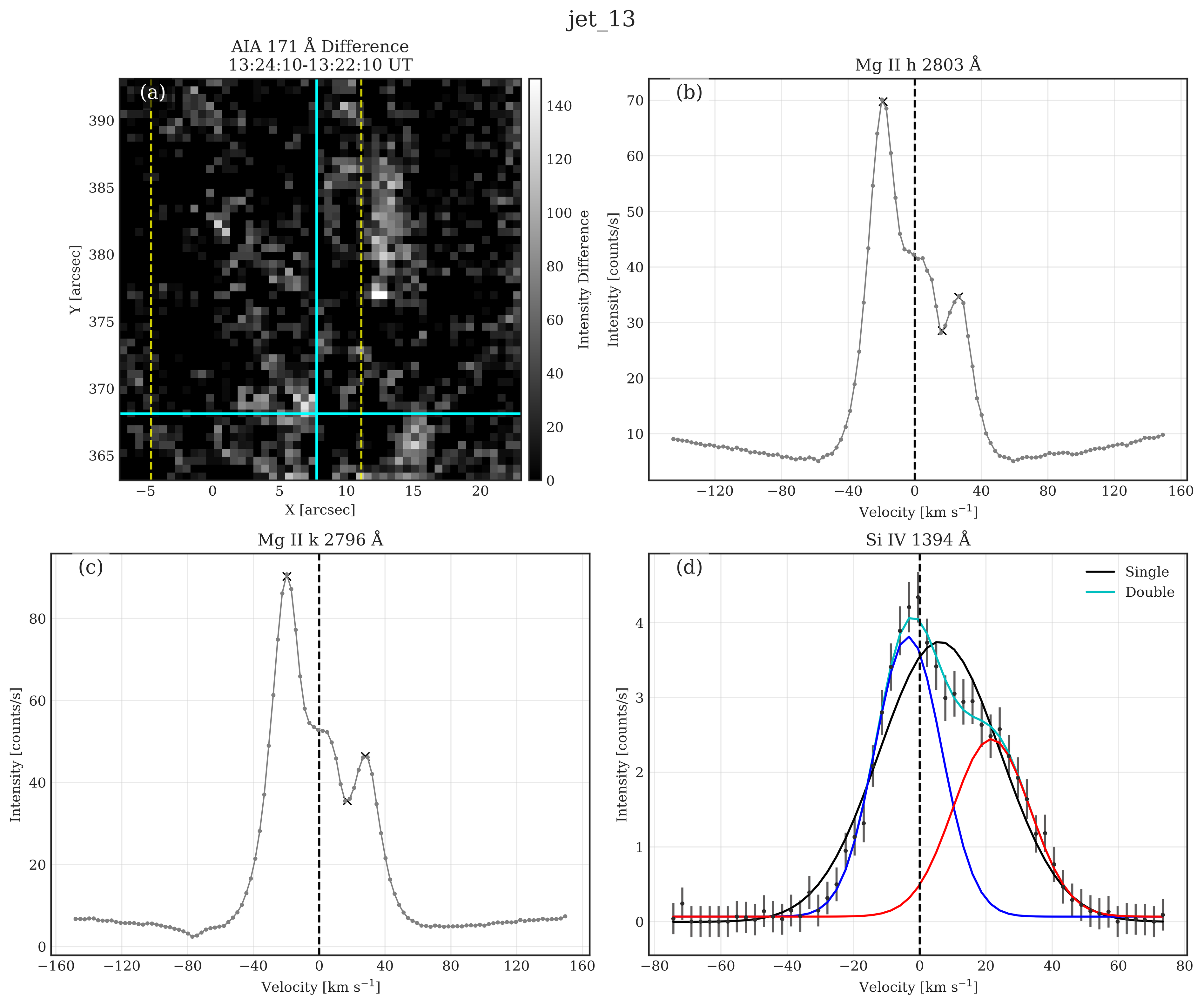}
    \caption{Same as Fig.~\ref{fig:jet06_difference_image} but for jet \_13.}
    \label{fig:jet_13_of_all_combined_images}
\end{figure}

We show that line profiles for the jet-13 in the  Fig.~\ref{fig:jet_13_of_all_combined_images}. We observe that the \ion{Mg}{2} line profiles for both the jet is quite peculiar. There appears a plateau just before the central dip in both h~\&~k lines.

\section{Summary and Discussions}\label{discussions}
In this work, we have investigated the chromospheric and TR counterparts of coronal plume footpoint activity. We combine SDO/AIA 171~{\AA} observations with IRIS slit-jaw images at 1400~{\AA} and \ion{Si}{4} and \ion{Mg}{2} spectral lines to perform a multi-thermal analysis spanning the middle chromosphere to the upper TR.

We identified rapid upflows at the base of the plumes from the temporal evolution of AIA 171~{\AA} images and examined the \ion{Si}{4} spectral lines. We defined an event as a jet when (1) a clear, collimated upflow was detected in AIA 171~{\AA} and (2) a blueshift was present in the secondary component of a double-Gaussian fit to the \ion{Si}{4} profile at the corresponding footpoint location. Although small-scale flows are ubiquitous in these regions, many were either weak or spectrally ambiguous; therefore, we restricted our analysis to events that satisfied both criteria robustly. Using this definition, we identified 19 jets suitable for detailed investigation. This approach is consistent with earlier studies that interpret strongly blueshifted secondary TR components as signatures of dynamic ejection events superposed on a relatively stationary background plasma \citep{koletti-2024, hosseini-2024}.

For most of the jets, the measured FWHM range between 20{--}50~km~s$^{-1}$, much larger than the thermal velocity of \ion{Si}{4} ($\approx$6--7~km~s$^{-1}$). This implies substantial non-thermal broadening at the footpoints of the jets. This can be interpreted as evidence for unresolved substructure, such as multi-threaded outflows, turbulence, or line-of-sight velocity gradients. The co-spatial and co-temporal presence of blueshifts with significantly enhanced non-thermal width is suggestive of impulsive energy release, which is consistent with the results obtained by \citep{pant-2014, hosseini-2024, malaker-2024}, who attribute the formation of jet like transients to magnetic reconnection at the base of coronal plumes. \citet{panesar-2018} reported significant flux cancellations prior to the eruption of mini-filaments, which may act as precursors to network jets through magnetic reconnection. However, we note that while the base of the plume was located in strong concentration of magnetic flux, we did not observe any noticeable change. Such changes may be below the current detection limit of magnetic field measurements, as was also reported by \citep{cho-2023,wang-2016}.

We simultaneously observe the \ion{Mg}{2}~h and k lines, which probe the middle-to-upper chromosphere and reveal highly complex response. While, signature of both blueshifts and redshifts are present, \ion{Mg}{2} exhibits predominant redshifts rather than blueshifts at the site of these jets. The correlation studies clearly demonstrate that while there is no correlation between the red component of \ion{Si}{4} and chromospheric downflows (Fig.~\ref{fig:jointplot_mgk_vs_si}(a), top row), there is somewhat pronounced anti-correlation in TR upflows and chromospheric upflows (Fig.~\ref{fig:jointplot_mgk_vs_si}(a) bottom row). Moreover, there is a strong correlation between TR and chromospheric downflows. These correlations are similar to those obtained for CHs and QS by \cite{Upendran_2021_C2, upendran_2022_Mg2}, where correlated TR upflows with chromospheric downflows are strongly suggestive of reconnection occurring at the base of TR. Such reconnection events drive bidirectional flows in the atmosphere seen in TR and chromospheric lines. However, given the absence of any significant minority polarity in the plume, the plume formation and sustenance may occur due to flux cancellation through convergence~\citep{avallone_plumes}, or potentially through small angle reconnection~\citep{nanojet}.

We note that jets and spicules produced through ambipolar diffusion in the partially ionized chromosphere \citep{martnezsykora-2017,martnezsykora-2020} can also exhibit strong blueshifts and enhanced non-thermal widths, as can Alfv\'{e}nic waves ubiquitous in open coronal 
field regions \citep{morton-2015, morton-2023a}. However, neither mechanism naturally produces the observed pattern of correlated TR upflows and chromospheric downflows, which is a distinctive signature of interchange reconnection at the TR base. We therefore interpret the jets in our sample as predominantly reconnection-driven, while acknowledging that Alfv\'{e}nic waves may be present within the jets regardless of their formation mechanism. Albeit for different spectral lines (and reconnection at different heights), such a correlations, is also seen in simulations by \cite{Upendran_2025_sim}.

These results provides further inputs into the formation of plumes and may be used to constrain simulations dealing with mass and energy transport in dynamically coupled magnetized solar atmosphere. We note that similar energizing mechanisms have been reported at the footpoints of various coronal loop systems. \citep{bryans-2016,martnezsykora-2018,chaurasiya-2024}, with their manifestation shaped by the overarching magnetic topology. \citet{bose-2025} investigated spicular activity at quiet-Sun and coronal hole regions using IRIS and AIA DEM analysis, finding blueshifts in both \ion{Mg}{2} ($20$--$30$~km~s$^{-1}$) and \ion{Si}{4} ($12$--$20$~km~s$^{-1}$) consistent with spicular ejection heated to upper-TR/lower-coronal temperatures ($\log T[\mathrm{K}] = 5.7$--$5.9$). Some of the jets studied here may include such spicular activity; however, while \citet{bose-2025} focused on the general coronal signature of spicules, we specifically target coronal hole plume footpoints, where the additional structure and long-term sustenance of the plume itself is of interest. We identify collimated coronal jets in AIA 171~\AA{}, co-spatial and co-temporal (within $\sim$15~s, comparable to the spectral exposure time) with the TR and chromospheric response, and find \ion{Si}{4} upflows paired with \ion{Mg}{2} downflows, a signature we interpret as evidence of interchange reconnection.  However, further work using multi-spectral line analysis is required to fully comprehend the chromospheric-corona connection in plumes, which will be possible with future observations taken by the upcoming missions such as MUlti-slit Solar Explorer \citep[][]{muse} and Solar-C \citep[][]{solarc}.

\begin{acknowledgements}
    We thank the referee for the constructive feedback on the manuscript. We acknowledge the CEFIPRA grant 6904-2. U.V. acknowledges support by NASA contracts NNG09FA40C (IRIS), 80GSFC21C0011 (MUSE), and NASA grant 80NSSC25K7956 (HGI). We gratefully acknowledge the insightful discussions with Late Prof Rob Rutten during early phase of this study. IRIS is a NASA small explorer mission developed and operated by LMSAL with mission operations executed at NASA Ames Research Center and major contributions to downlink communications funded by ESA and the Norwegian Space Centre. We acknowledge the use of data from AIA and HMI. AIA and HMI are instruments on board SDO, a mission for NASA’s Living With a Star program. 
\end{acknowledgements}

\appendix \label{sec:appendix}
\restartappendixnumbering
\section{Spectral line profiles for all the remaining jets}

Spectral lines and difference images for all the remaining jets, except jet-06, jet-13, and jet-18, which are discussed in the main text.

\begin{figure}[p]
    \centering
    \begin{subfigure}{0.48\textwidth}
        \includegraphics[width=\textwidth]{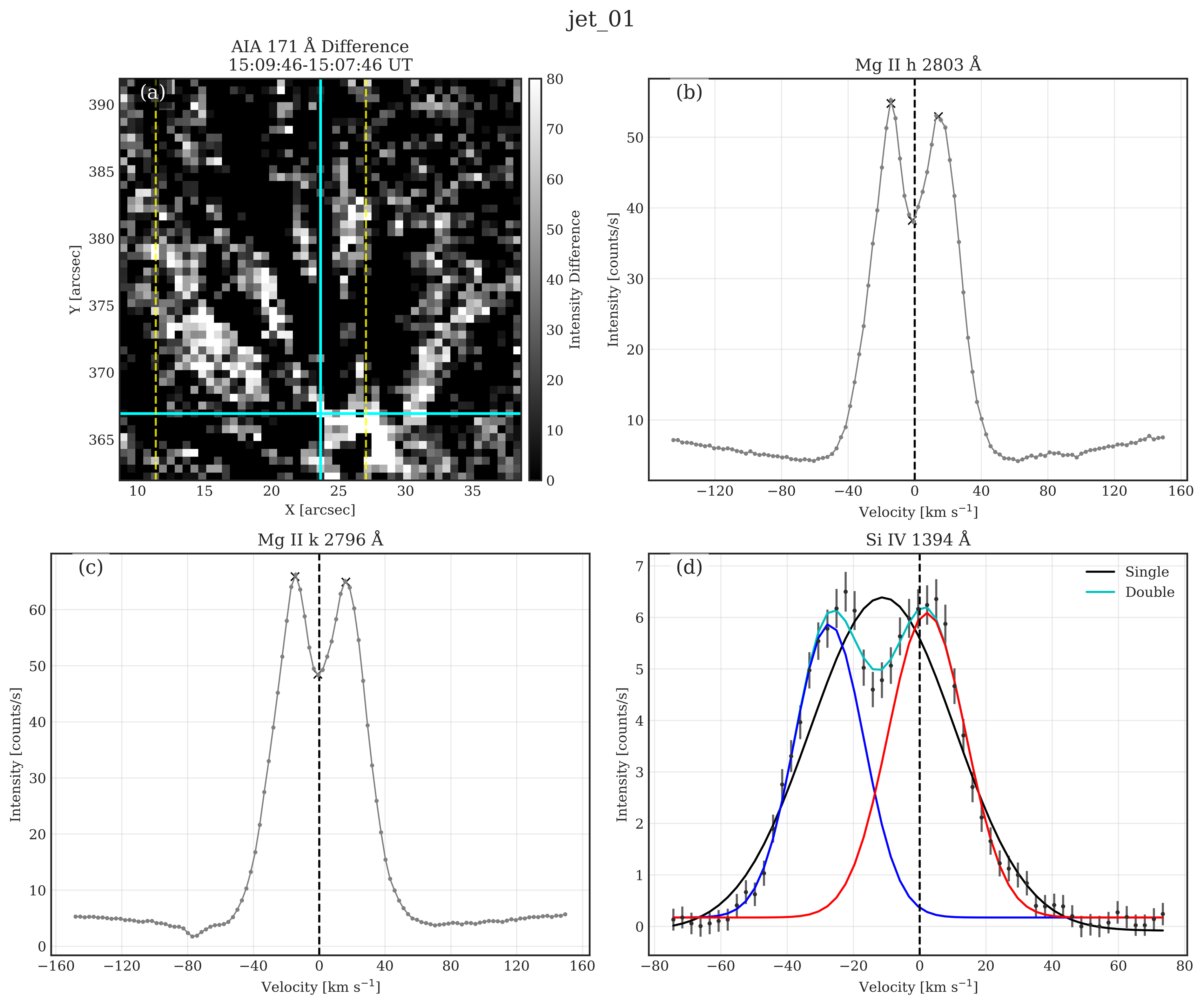}
        \caption{jet\_01}
    \end{subfigure}
    \hfill
    \begin{subfigure}{0.48\textwidth}
        \includegraphics[width=\textwidth]{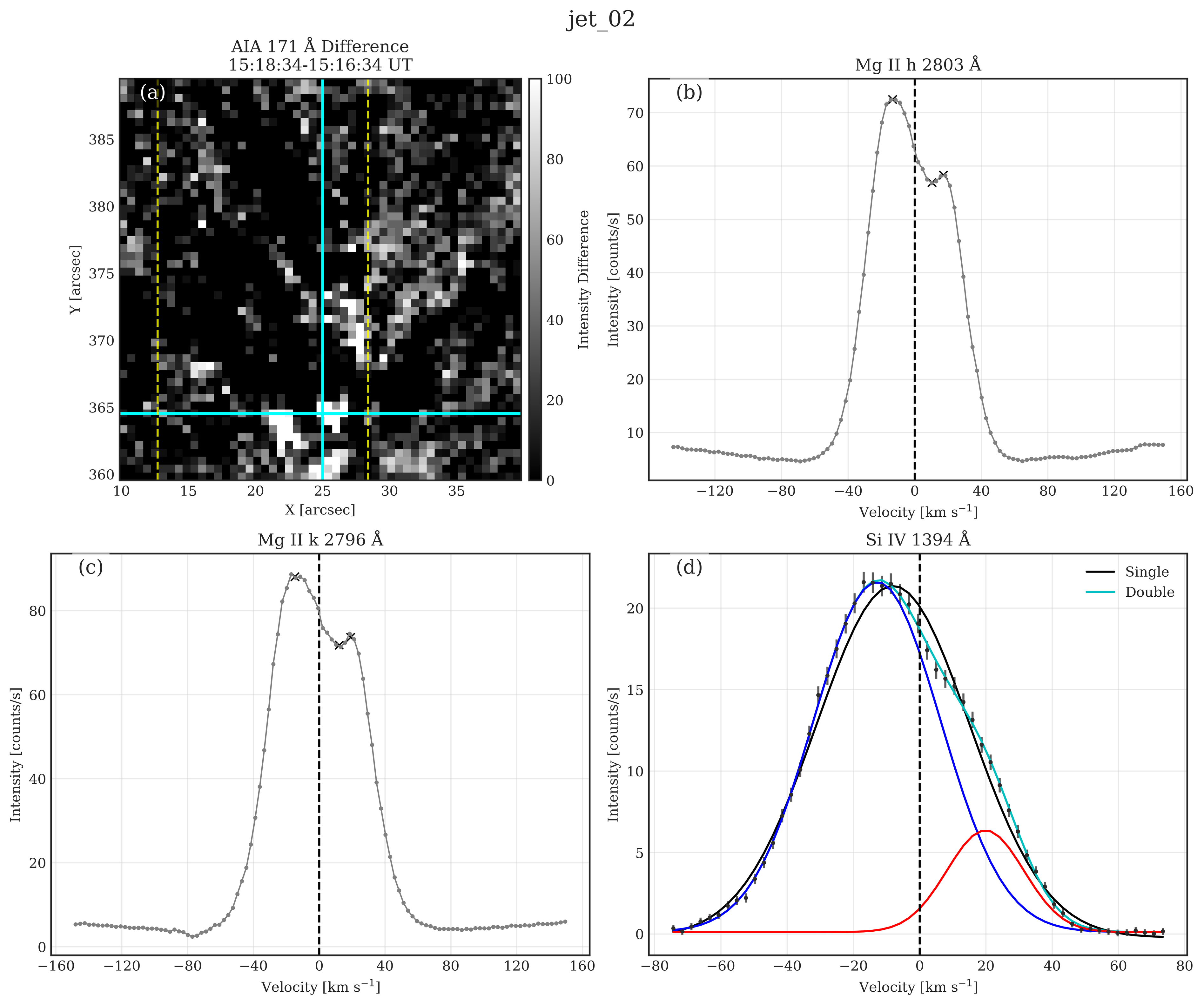}
        \caption{jet\_02}
    \end{subfigure}

    \vspace{1em}

    \begin{subfigure}{0.48\textwidth}
        \includegraphics[width=\textwidth]{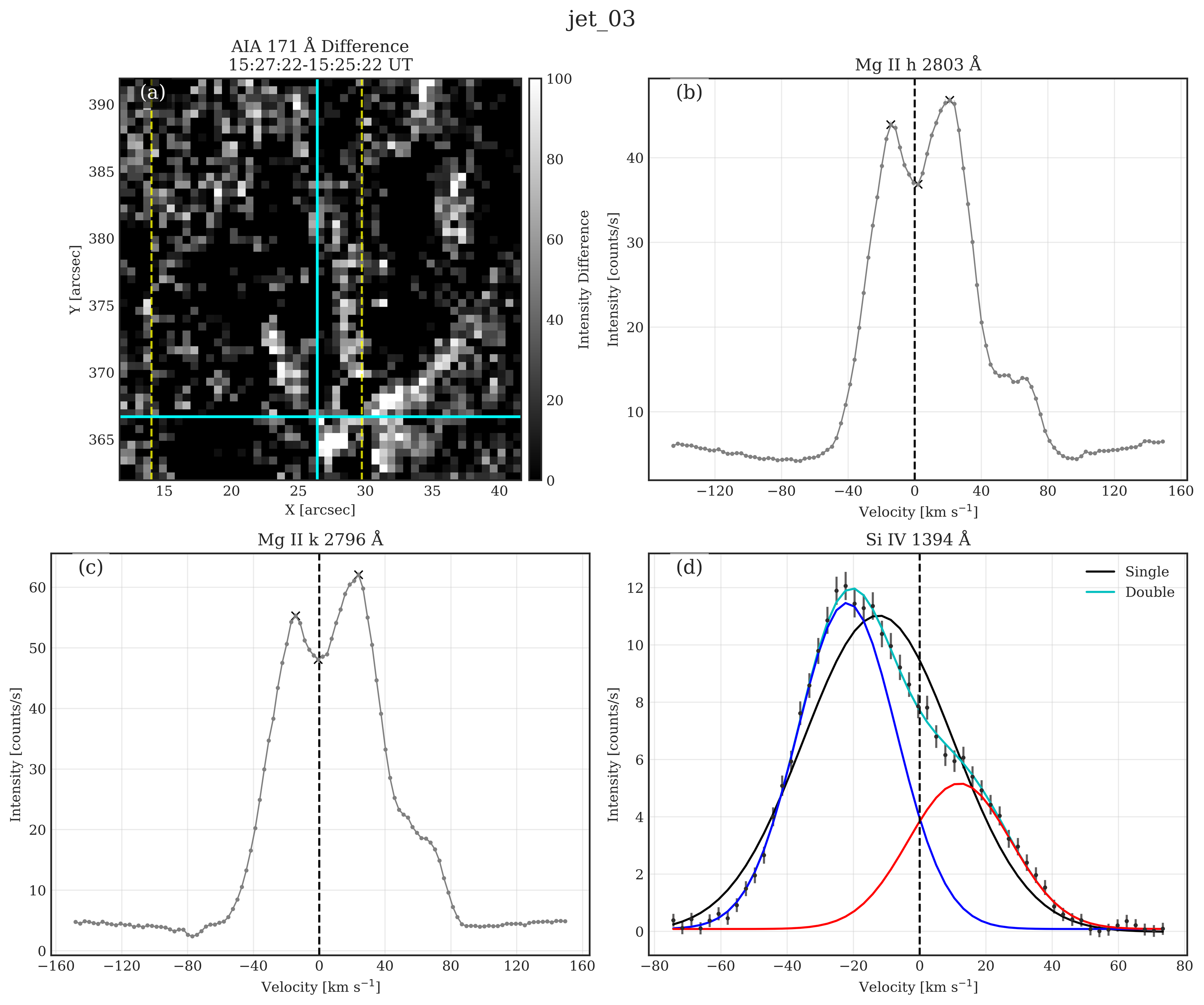}
        \caption{jet\_03}
    \end{subfigure}
    \hfill
    \begin{subfigure}{0.48\textwidth}
        \includegraphics[width=\textwidth]{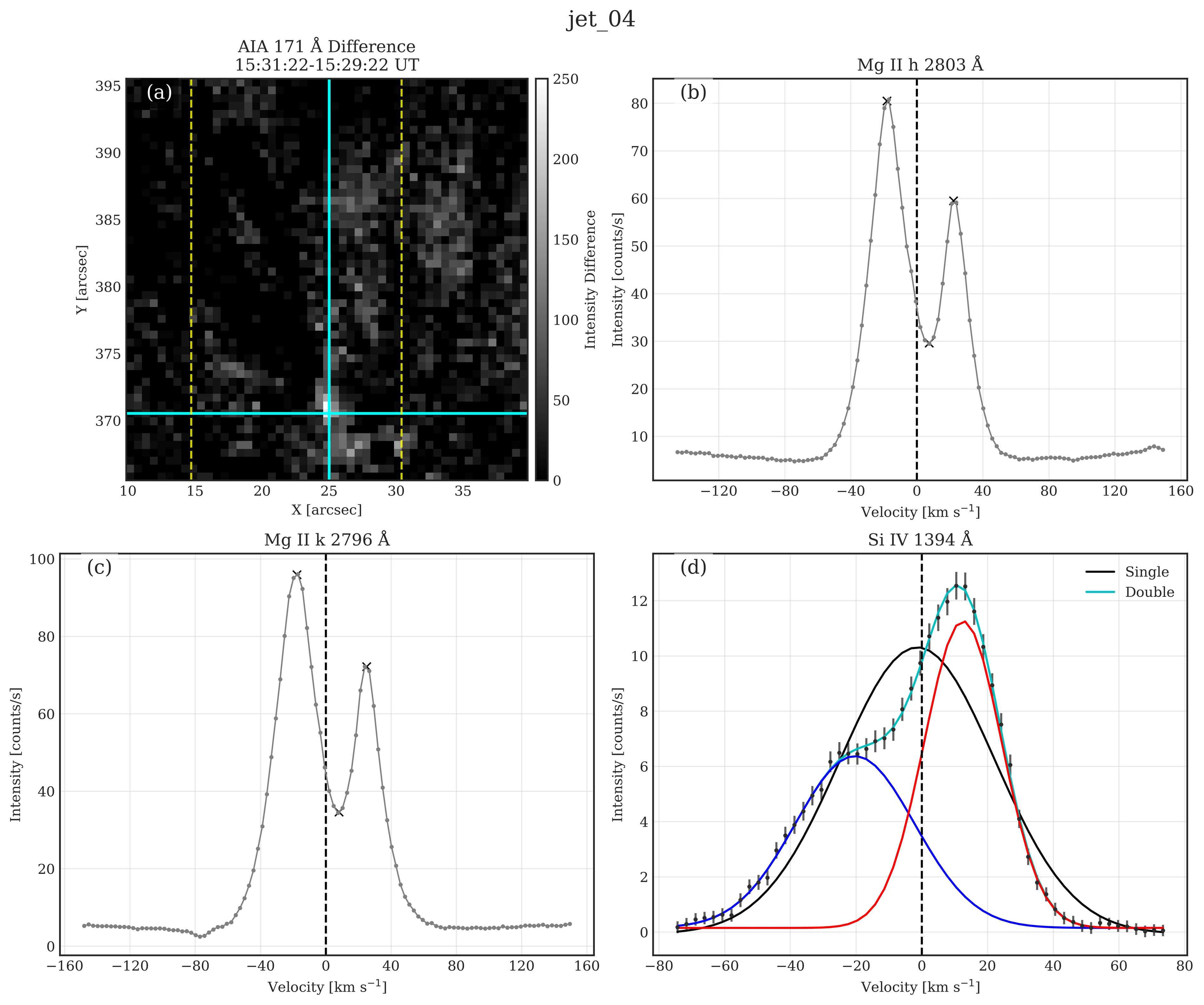}
        \caption{jet\_04}
    \end{subfigure}

    \vspace{1em}

    \begin{subfigure}{0.48\textwidth}
        \includegraphics[width=\textwidth]{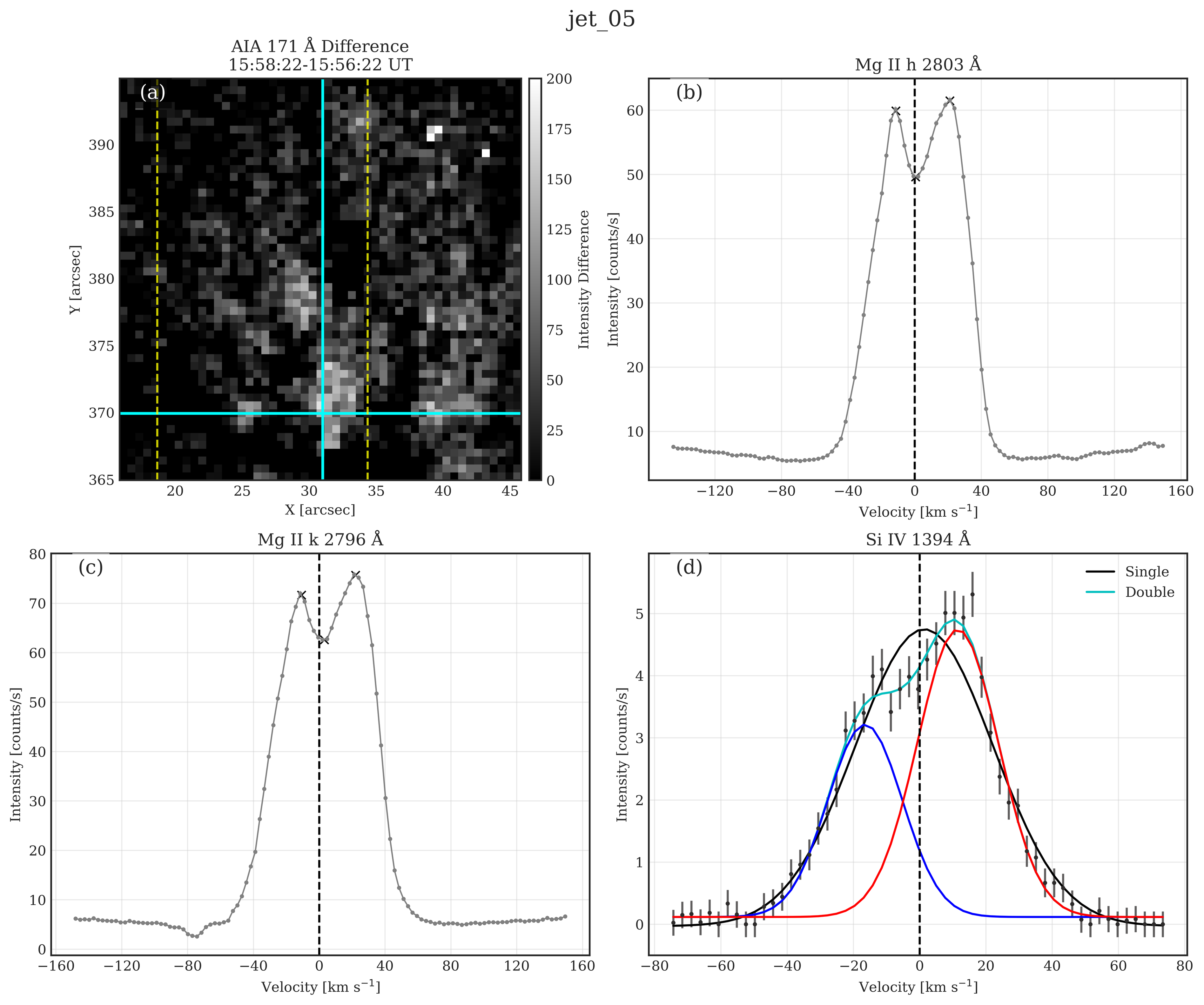}
        \caption{jet\_05}
    \end{subfigure}
    \hfill
    \begin{subfigure}{0.48\textwidth}
        \includegraphics[width=\textwidth]{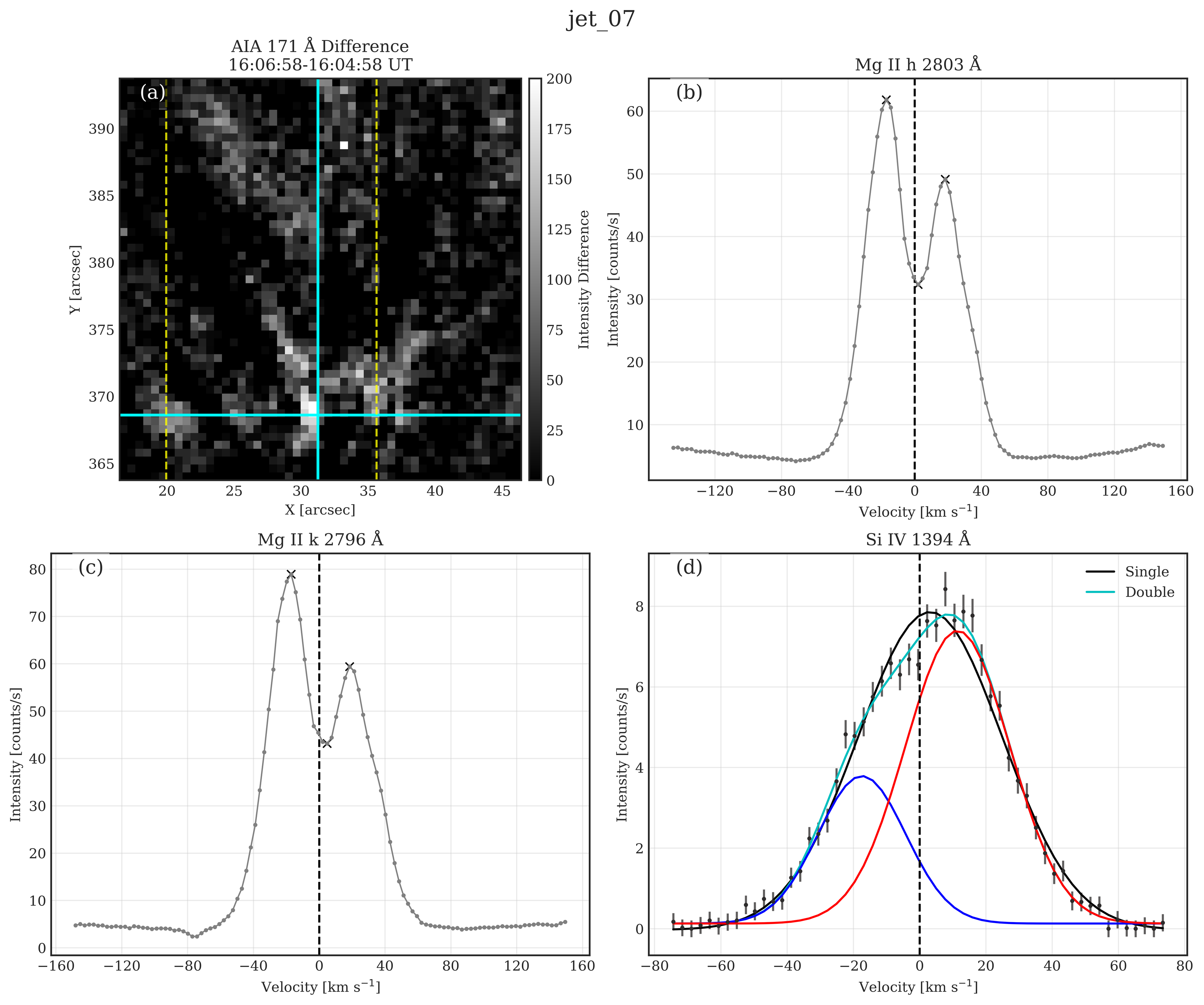}
        \caption{jet\_07}
    \end{subfigure}
    \caption{AIA difference images and spectral lines labeled from jets 01--07.}
    \label{fig:jets_A1}
\end{figure}

\begin{figure}[p]
    \centering
    \begin{subfigure}{0.48\textwidth}
        \includegraphics[width=\textwidth]{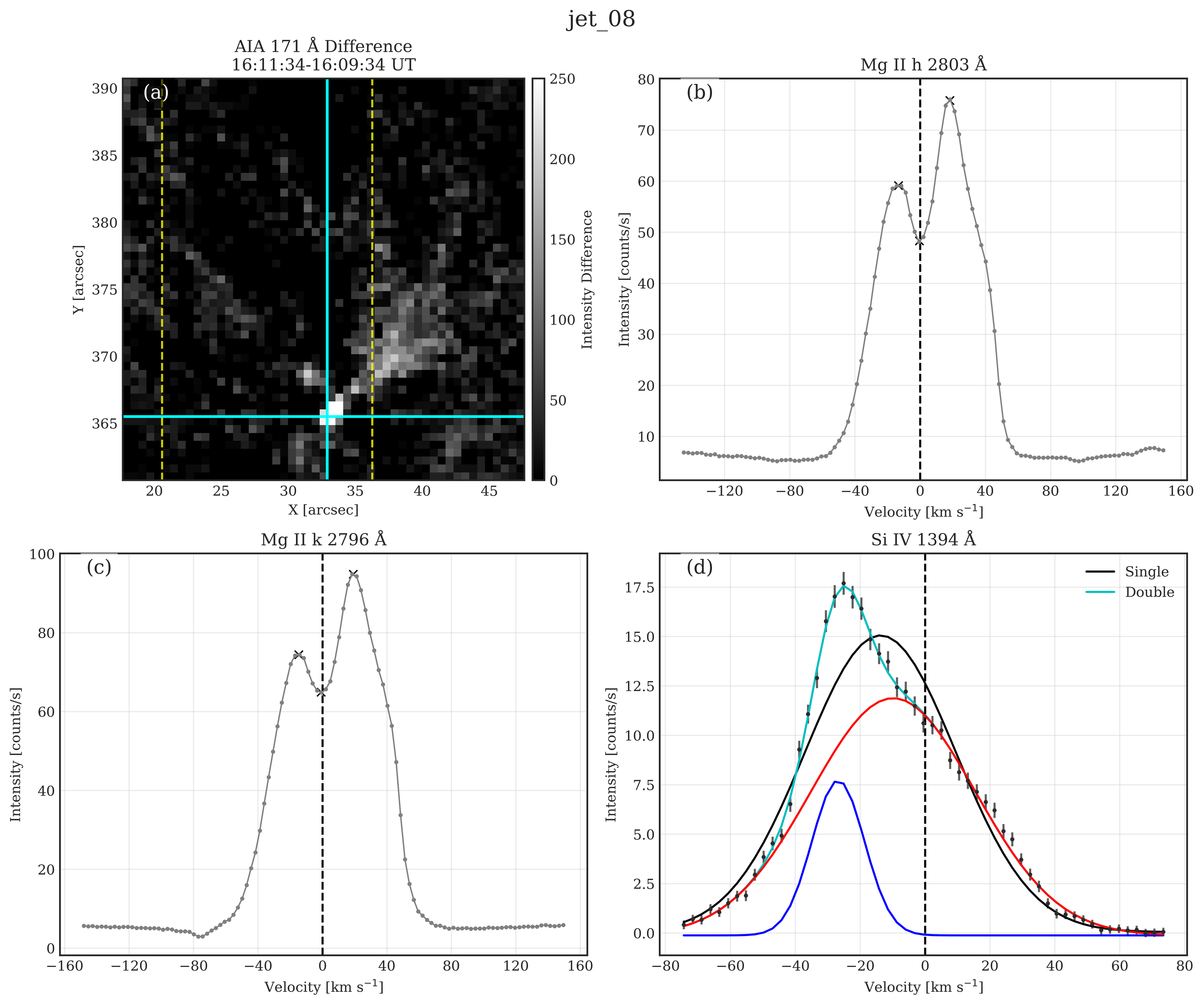}
        \caption{jet\_08}
    \end{subfigure}
    \hfill
    \begin{subfigure}{0.48\textwidth}
        \includegraphics[width=\textwidth]{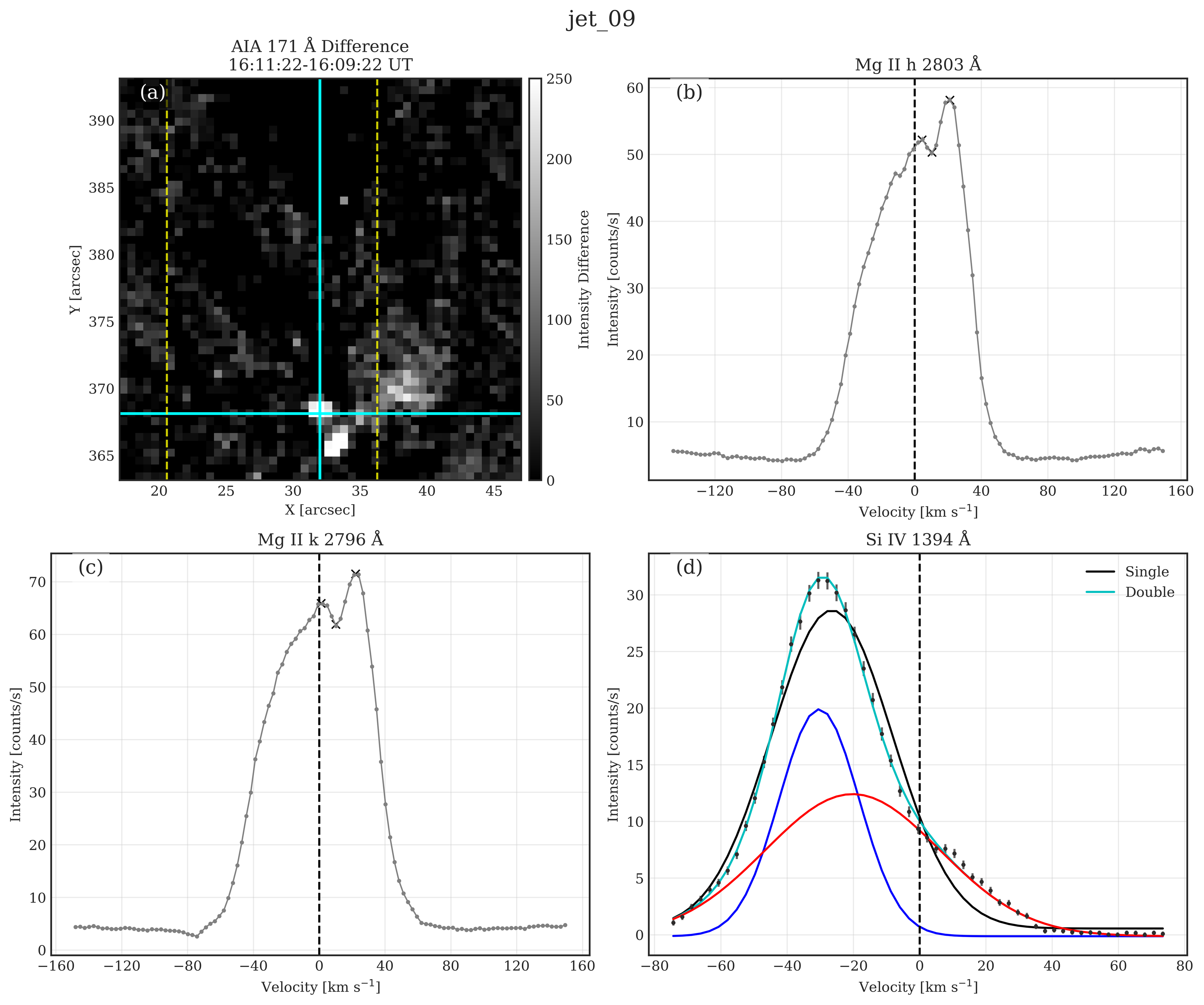}
        \caption{jet\_09}
    \end{subfigure}

    \vspace{1em}

    \begin{subfigure}{0.48\textwidth}
        \includegraphics[width=\textwidth]{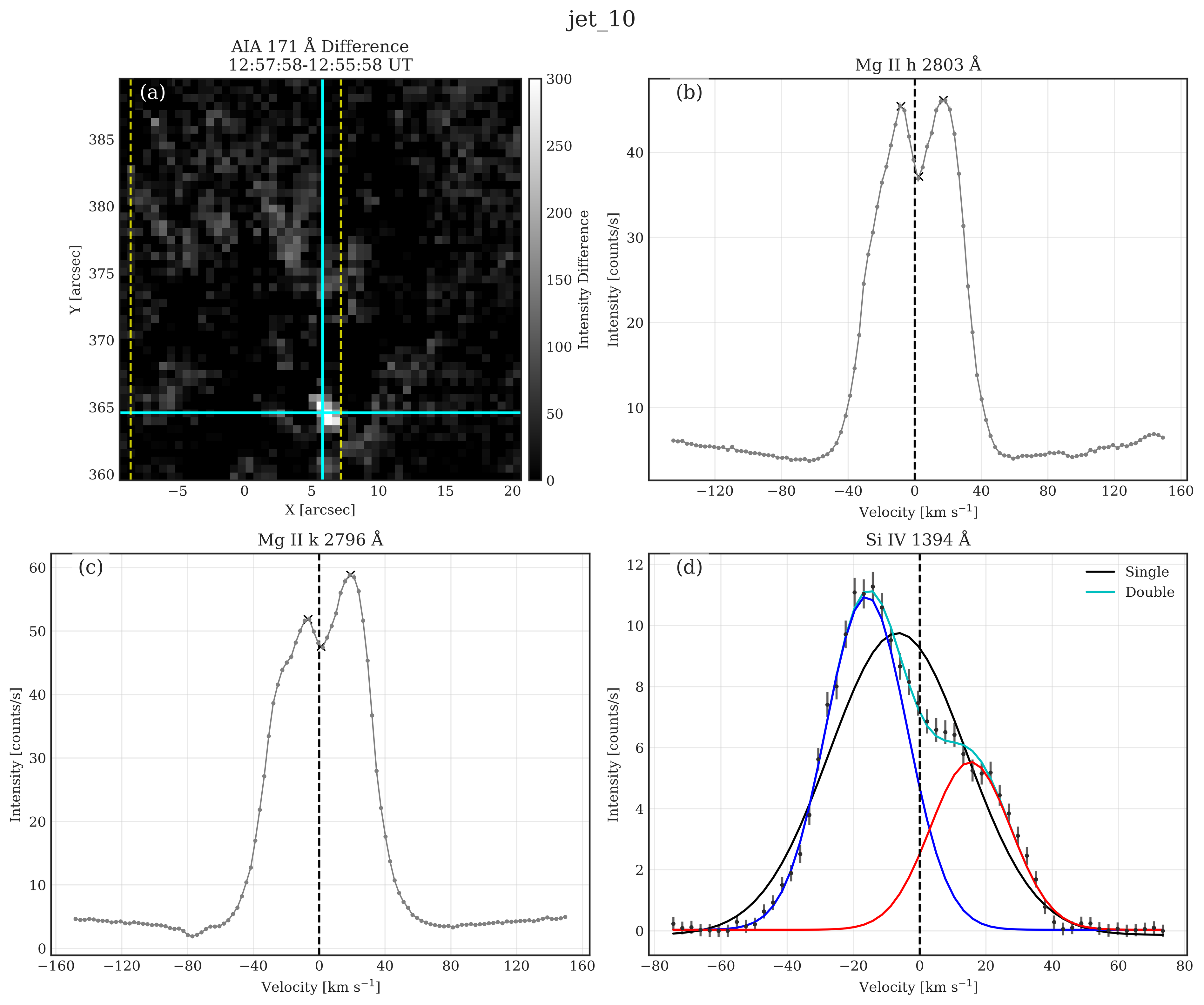}
        \caption{jet\_10}
    \end{subfigure}
    \hfill
    \begin{subfigure}{0.48\textwidth}
        \includegraphics[width=\textwidth]{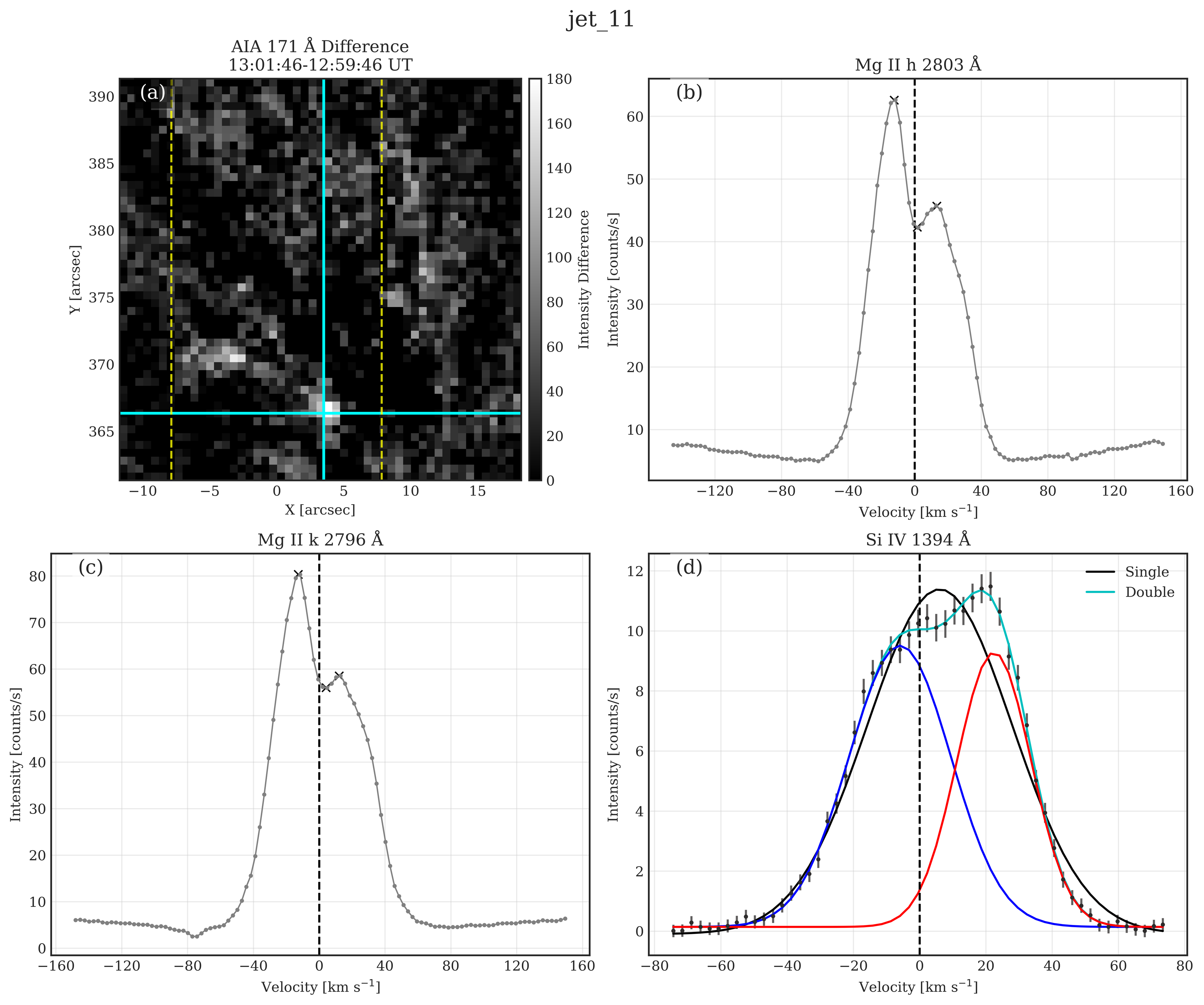}
        \caption{jet\_11}
    \end{subfigure}

    \vspace{1em}

    \begin{subfigure}{0.48\textwidth}
        \includegraphics[width=\textwidth]{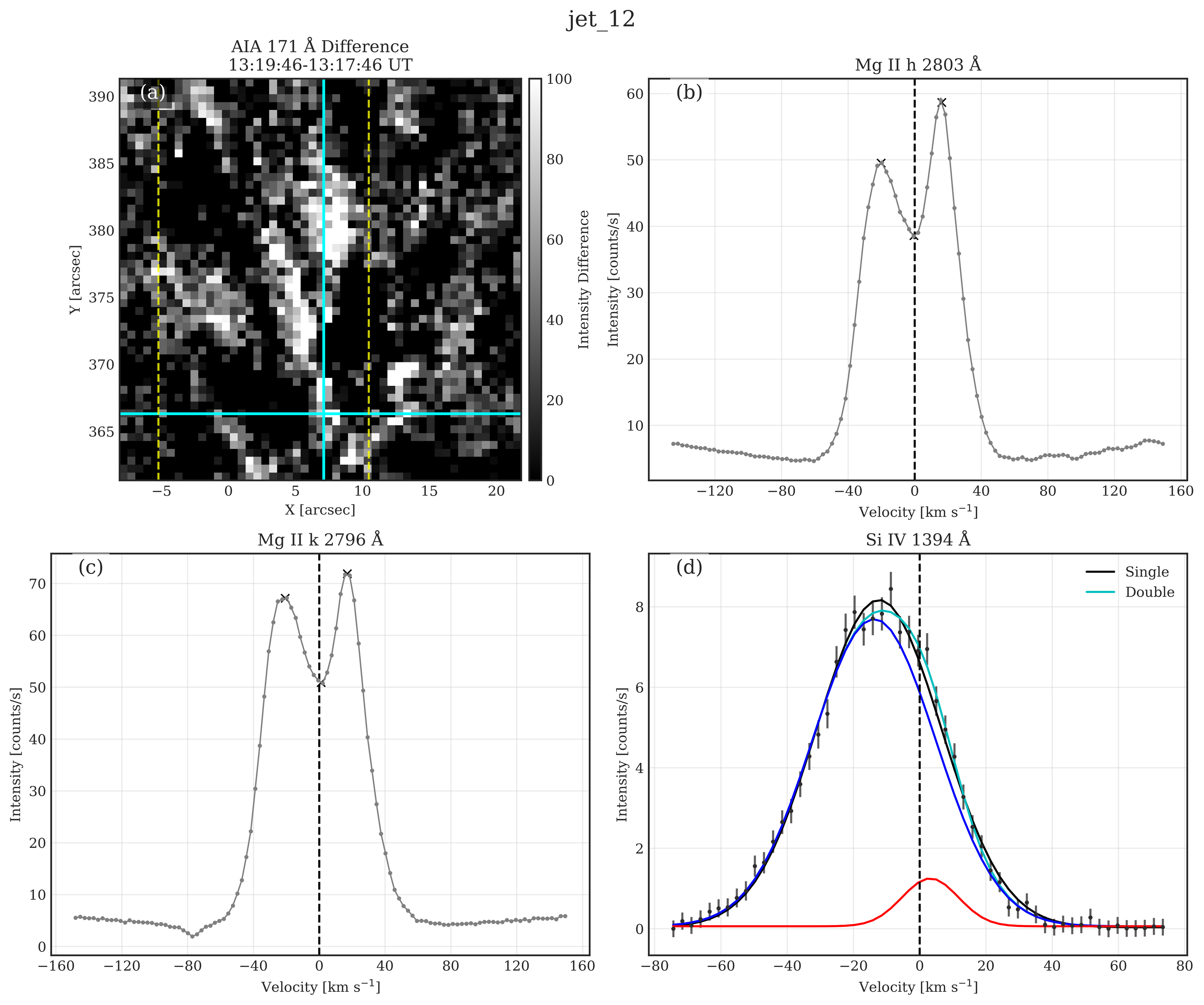}
        \caption{jet\_12}
    \end{subfigure}
    \hfill
    \begin{subfigure}{0.48\textwidth}
        \includegraphics[width=\textwidth]{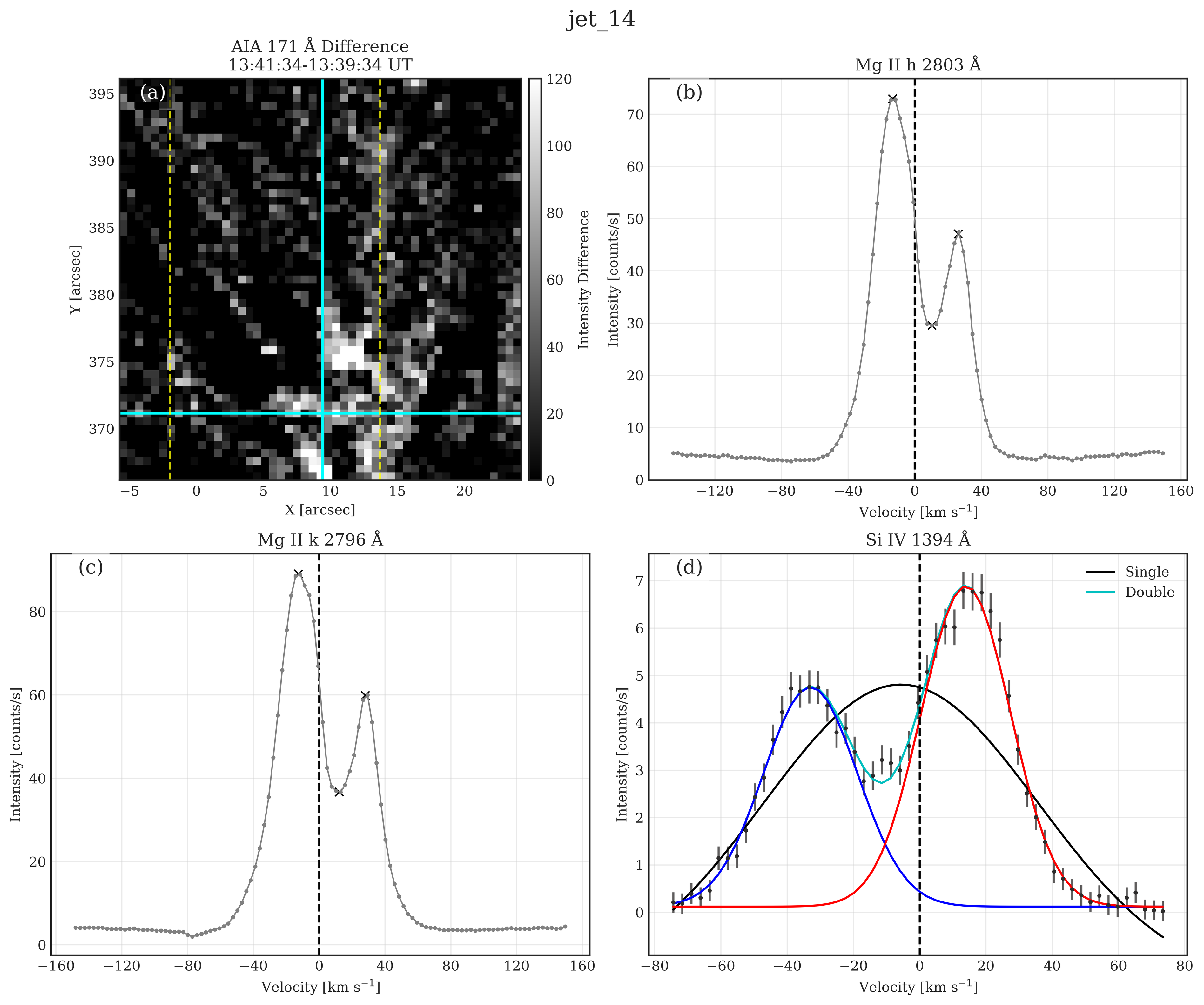}
        \caption{jet\_14}
    \end{subfigure}
    \caption{AIA difference images and spectral lines labeled from jets 08--14.}
    \label{fig:jets_A2}

\end{figure}

\begin{figure}[p]
    \centering
    \begin{subfigure}{0.48\textwidth}
        \includegraphics[width=\textwidth]{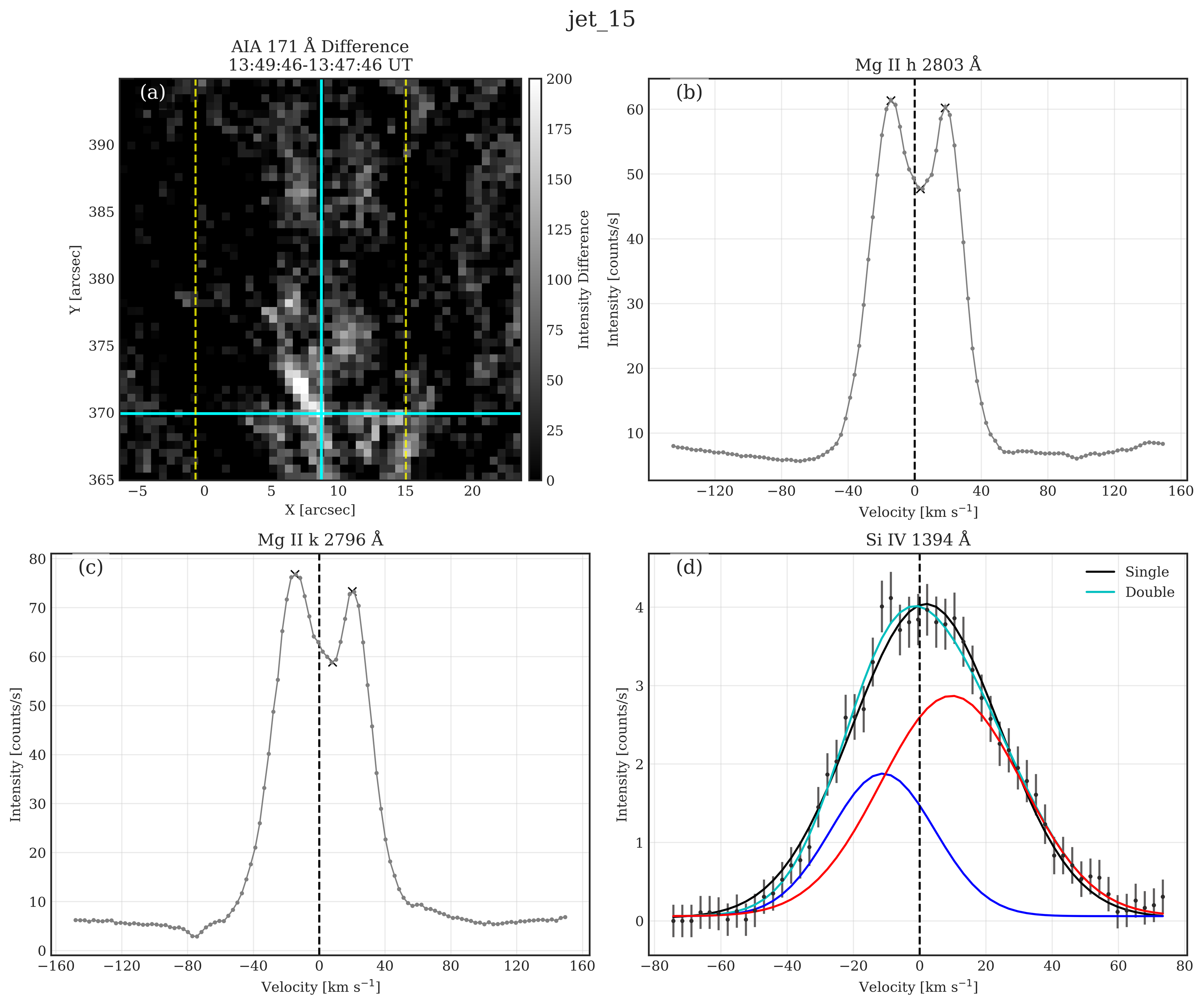}
        \caption{jet\_15}
    \end{subfigure}
    \hfill
    \begin{subfigure}{0.48\textwidth}
        \includegraphics[width=\textwidth]{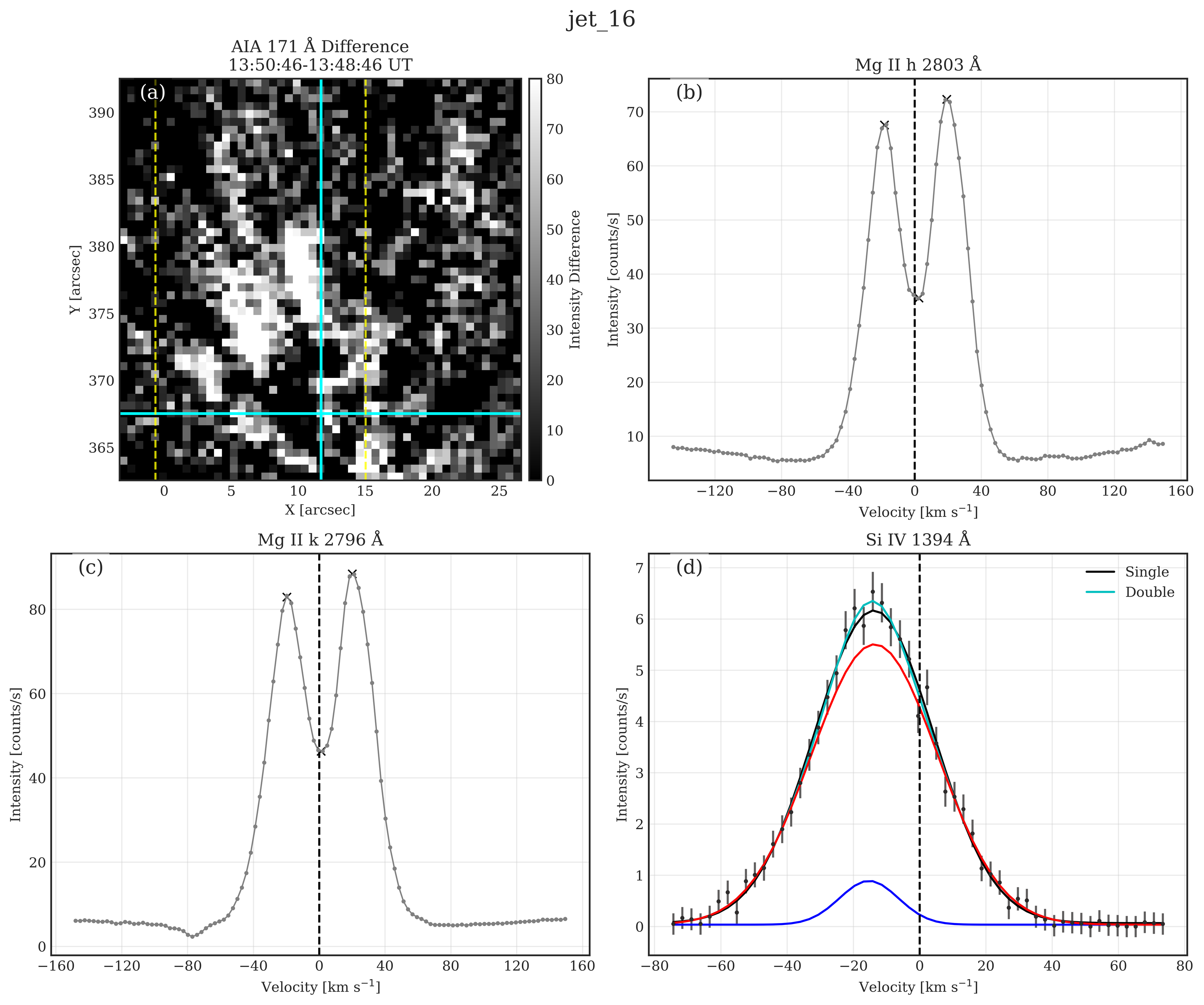}
        \caption{jet\_16}
    \end{subfigure}

    \vspace{1em}

    \begin{subfigure}{0.48\textwidth}
        \includegraphics[width=\textwidth]{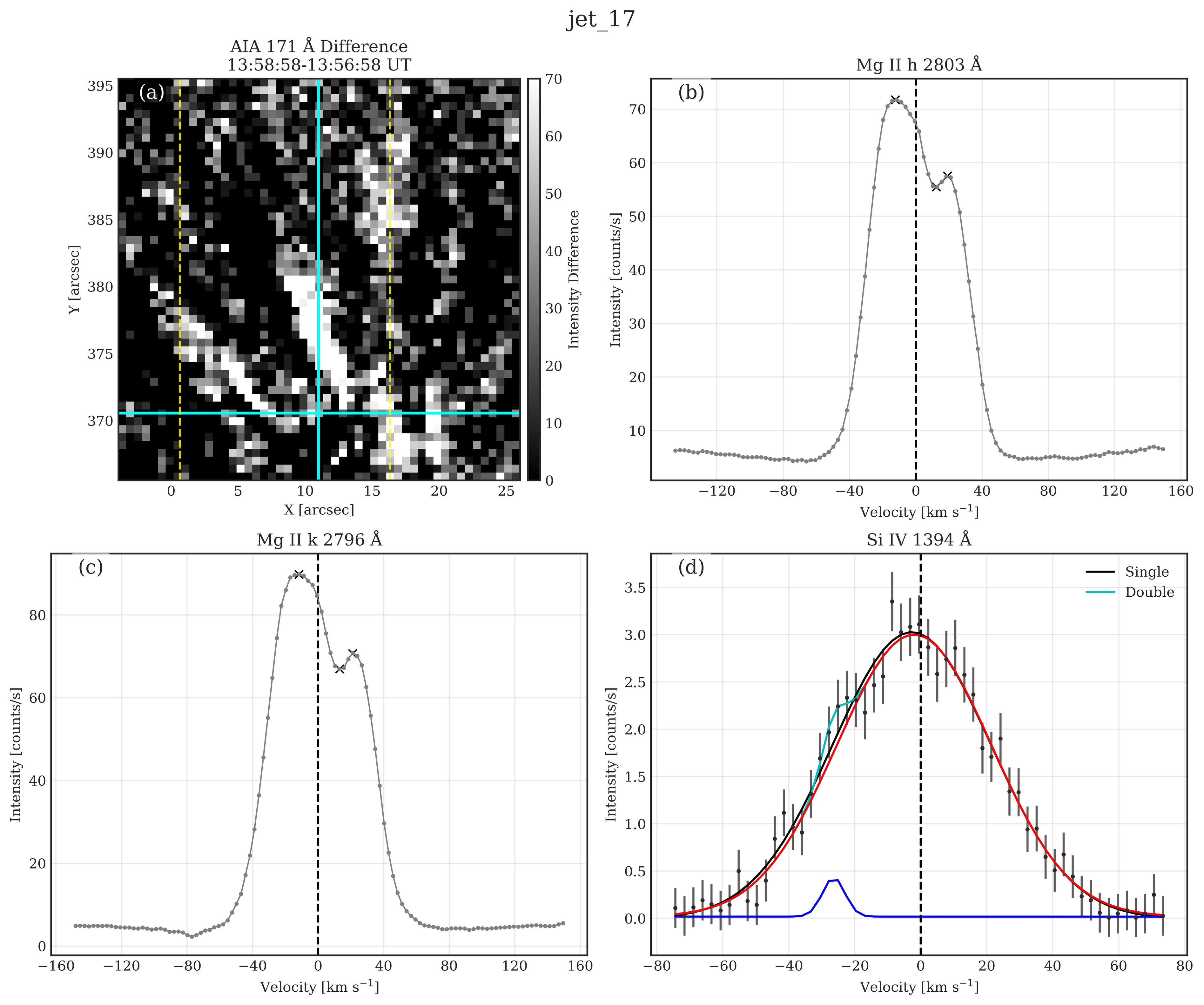}
        \caption{jet\_17}
    \end{subfigure}
    \hfill
    \begin{subfigure}{0.48\textwidth}
        \includegraphics[width=\textwidth]{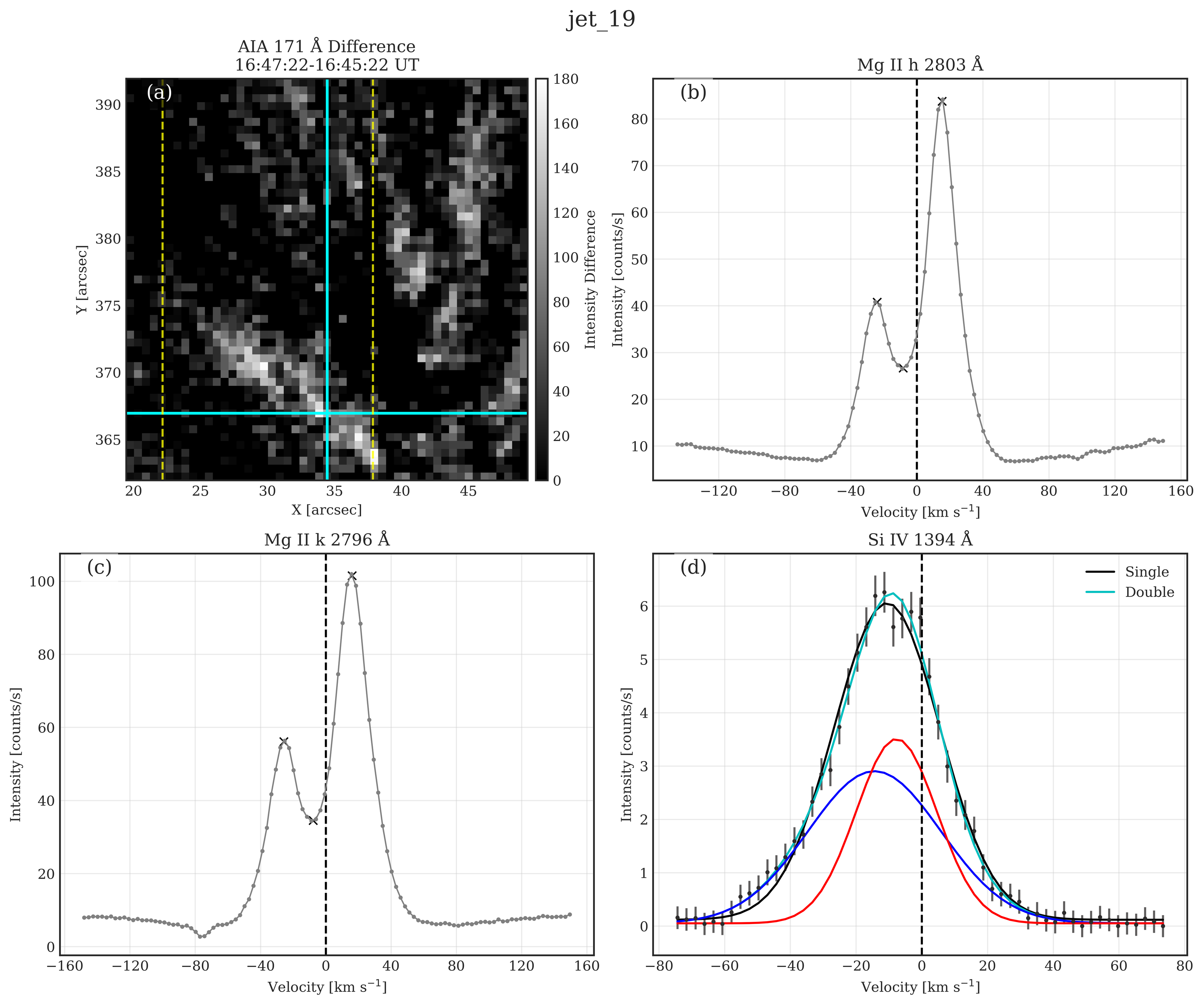}
        \caption{jet\_19}
    \end{subfigure}

    \caption{AIA difference images and spectral lines labeled from jets 15--19.}
    \label{fig:jets_A3}
    \vspace{1em}

\end{figure}


\begin{thebibliography}{}
\expandafter\ifx\csname natexlab\endcsname\relax\def\natexlab#1{#1}\fi
\providecommand{\url}[1]{\href{#1}{#1}}
\providecommand{\dodoi}[1]{doi:~\href{http://doi.org/#1}{\nolinkurl{#1}}}
\providecommand{\doeprint}[1]{\href{http://ascl.net/#1}{\nolinkurl{http://ascl.net/#1}}}
\providecommand{\doarXiv}[1]{\href{https://arxiv.org/abs/#1}{\nolinkurl{https://arxiv.org/abs/#1}}}

% type= article
\bibitem[{I.~A. {Ahmad} \& D.~F. {Webb}(1978){Ahmad} \& {Webb}}]{ahmad-1978}
{Ahmad}, I.~A., \& {Webb}, D.~F. 1978, \bibinfo{title}{{X-ray analysis of a
  polar plume.},} \solphys, 58, 323, \dodoi{10.1007/BF00157278}

% type= article
\bibitem[{I.~A. {Ahmad} \& G.~L. {Withbroe}(1977){Ahmad} \&
  {Withbroe}}]{ahmad-1977}
{Ahmad}, I.~A., \& {Withbroe}, G.~L. 1977, \bibinfo{title}{{EUV analysis of
  polar plumes.},} \solphys, 53, 397, \dodoi{10.1007/BF00160283}

% type= article
\bibitem[{M.~J. {Allen} {et~al.}(1997){Allen}, {Oluseyi}, {Walker}, {Hoover},
  \& {Barbee}}]{allen-1997}
{Allen}, M.~J., {Oluseyi}, H.~M., {Walker}, A. B.~C., {Hoover}, R.~B., \&
  {Barbee}, Jr., T.~W. 1997, \bibinfo{title}{{CHROMOSPHERIC AND CORONAL
  STRUCTURE OF POLAR PLUMES - I. Magnetic Structure and Radiative Energy
  Balance},} \solphys, 174, 367, \dodoi{10.1023/A:1004955129119}

% type= article
\bibitem[{P. {Antolin} {et~al.}(2021){Antolin}, {Pagano}, {Testa}, {Petralia},
  \& {Reale}}]{nanojet}
{Antolin}, P., {Pagano}, P., {Testa}, P., {Petralia}, A., \& {Reale}, F. 2021,
  \bibinfo{title}{{Reconnection nanojets in the solar corona},} Nature
  Astronomy, 5, 54, \dodoi{10.1038/s41550-020-1199-8}

% type= article
\bibitem[{E.~A. {Avallone} {et~al.}(2018){Avallone}, {Tiwari}, {Panesar},
  {Moore}, \& {Winebarger}}]{avallone_plumes}
{Avallone}, E.~A., {Tiwari}, S.~K., {Panesar}, N.~K., {Moore}, R.~L., \&
  {Winebarger}, A. 2018, \bibinfo{title}{{Critical Magnetic Field Strengths for
  Solar Coronal Plumes in Quiet Regions and Coronal Holes?},} \apj, 861, 111,
  \dodoi{10.3847/1538-4357/aac82c}

% type= article
\bibitem[{S. {Bose} {et~al.}(2025){Bose}, {Joshi}, {Testa}, \& {De
  Pontieu}}]{bose-2025}
{Bose}, S., {Joshi}, J., {Testa}, P., \& {De Pontieu}, B. 2025,
  \bibinfo{title}{{On the Million-degree Signature of Spicules},} \apjl, 983,
  L7, \dodoi{10.3847/2041-8213/adc30d}

% type= article
\bibitem[{M.~A. {Branch} {et~al.}(1999){Branch}, {Coleman}, \&
  {Li}}]{branch-1999}
{Branch}, M.~A., {Coleman}, T.~F., \& {Li}, Y. 1999, \bibinfo{title}{{A
  Subspace, Interior, and Conjugate Gradient Method for Large-Scale
  Bound-Constrained Minimization Problems},} SIAM Journal on Scientific
  Computing, 21, 1, \dodoi{10.1137/S1064827595289108}

% type= article
\bibitem[{P. {Bryans} {et~al.}(2016){Bryans}, {McIntosh}, {De Moortel}, \& {De
  Pontieu}}]{bryans-2016}
{Bryans}, P., {McIntosh}, S.~W., {De Moortel}, I., \& {De Pontieu}, B. 2016,
  \bibinfo{title}{{On the Connection between Propagating Solar Coronal
  Disturbances and Chromospheric Footpoints},} \apjl, 829, L18,
  \dodoi{10.3847/2041-8205/829/1/L18}

% type= article
\bibitem[{R. {Chaurasiya} {et~al.}(2024){Chaurasiya}, {Bayanna}, {Louis},
  {Pereira}, \& {Mathew}}]{chaurasiya-2024}
{Chaurasiya}, R., {Bayanna}, A.~R., {Louis}, R.~E., {Pereira}, T.~M.~D., \&
  {Mathew}, S.~K. 2024, \bibinfo{title}{{On the Response of the Transition
  Region and the Corona to Rapid Excursions in the Chromosphere},} \apj, 970,
  179, \dodoi{10.3847/1538-4357/ad50d5}

% type= article
\bibitem[{M.~C.~M. {Cheung} {et~al.}(2022){Cheung}, {Mart{\'\i}nez-Sykora},
  {Testa}, {De Pontieu}, {Chintzoglou}, {Rempel}, {Polito}, {Kerr}, {Reeves},
  {Fletcher}, {Jin}, {N{\'o}brega-Siverio}, {Danilovic}, {Antolin}, {Allred},
  {Hansteen}, {Ugarte-Urra}, {DeLuca}, {Longcope}, {Takasao}, {DeRosa},
  {Boerner}, {Jaeggli}, {Nitta}, {Daw}, {Carlsson}, {Golub}, \& {The}}]{muse}
{Cheung}, M. C.~M., {Mart{\'\i}nez-Sykora}, J., {Testa}, P., {et~al.} 2022,
  \bibinfo{title}{{Probing the Physics of the Solar Atmosphere with the
  Multi-slit Solar Explorer (MUSE). II. Flares and Eruptions},} \apj, 926, 53,
  \dodoi{10.3847/1538-4357/ac4223}

% type= article
\bibitem[{K.-S. {Cho} {et~al.}(2021){Cho}, {Cho}, {Madjarska}, {Nakariakov},
  {Yang}, {Choi}, {Lim}, {Lee}, {Seough}, {Lee}, \& {Kim}}]{cho-2021}
{Cho}, K.-S., {Cho}, I.-H., {Madjarska}, M.~S., {et~al.} 2021,
  \bibinfo{title}{{On the Nature of Propagating Intensity Disturbances in Polar
  Plumes during the 2017 Total Solar Eclipse},} \apj, 909, 202,
  \dodoi{10.3847/1538-4357/abdfd5}

% type= article
\bibitem[{K.-S. {Cho} {et~al.}(2023){Cho}, {Kumar}, {Cho}, {Madjarska},
  {Nakariakov}, {Lim}, {Cao}, {Yurchyshyn}, {Yang}, \& {Park}}]{cho-2023}
{Cho}, K.-S., {Kumar}, P., {Cho}, I.-H., {et~al.} 2023,
  \bibinfo{title}{{High-resolution Observations of Plume Footpoints in a Solar
  Coronal Hole},} \apj, 953, 69, \dodoi{10.3847/1538-4357/acd456}

% type= article
\bibitem[{B. {De Pontieu} {et~al.}(2014){De Pontieu}, {Title}, {Lemen},
  {Kushner}, {Akin}, {Allard}, {Berger}, {Boerner}, {Cheung}, {Chou}, {Drake},
  {Duncan}, {Freeland}, {Heyman}, {Hoffman}, {Hurlburt}, {Lindgren}, {Mathur},
  {Rehse}, {Sabolish}, {Seguin}, {Schrijver}, {Tarbell}, {W{\"u}lser},
  {Wolfson}, {Yanari}, {Mudge}, {Nguyen-Phuc}, {Timmons}, {van Bezooijen},
  {Weingrod}, {Brookner}, {Butcher}, {Dougherty}, {Eder}, {Knagenhjelm},
  {Larsen}, {Mansir}, {Phan}, {Boyle}, {Cheimets}, {DeLuca}, {Golub}, {Gates},
  {Hertz}, {McKillop}, {Park}, {Perry}, {Podgorski}, {Reeves}, {Saar}, {Testa},
  {Tian}, {Weber}, {Dunn}, {Eccles}, {Jaeggli}, {Kankelborg}, {Mashburn},
  {Pust}, {Springer}, {Carvalho}, {Kleint}, {Marmie}, {Mazmanian}, {Pereira},
  {Sawyer}, {Strong}, {Worden}, {Carlsson}, {Hansteen}, {Leenaarts},
  {Wiesmann}, {Aloise}, {Chu}, {Bush}, {Scherrer}, {Brekke}, {Martinez-Sykora},
  {Lites}, {McIntosh}, {Uitenbroek}, {Okamoto}, {Gummin}, {Auker}, {Jerram},
  {Pool}, \& {Waltham}}]{iris-depontieu}
{De Pontieu}, B., {Title}, A.~M., {Lemen}, J.~R., {et~al.} 2014,
  \bibinfo{title}{{The Interface Region Imaging Spectrograph (IRIS)},}
  \solphys, 289, 2733, \dodoi{10.1007/s11207-014-0485-y}

% type= article
\bibitem[{C.~E. {DeForest} {et~al.}(1997){DeForest}, {Hoeksema}, {Gurman},
  {Thompson}, {Plunkett}, {Howard}, {Harrison}, \& {Hasslerz}}]{deforest-1997}
{DeForest}, C.~E., {Hoeksema}, J.~T., {Gurman}, J.~B., {et~al.} 1997,
  \bibinfo{title}{{Polar Plume Anatomy: Results of a Coordinated Observation},}
  \solphys, 175, 393, \dodoi{10.1023/A:1004955223306}

% type= article
\bibitem[{C.~E. {DeForest} {et~al.}(2001{\natexlab{a}}){DeForest}, {Lamy}, \&
  {Llebaria}}]{deforest-2001a}
{DeForest}, C.~E., {Lamy}, P.~L., \& {Llebaria}, A. 2001{\natexlab{a}},
  \bibinfo{title}{{Solar Polar Plume Lifetime and Coronal Hole Expansion:
  Determination from Long-Term Observations},} \apj, 560, 490,
  \dodoi{10.1086/322497}

% type= article
\bibitem[{C.~E. {DeForest} {et~al.}(2001{\natexlab{b}}){DeForest}, {Plunkett},
  \& {Andrews}}]{deforest-2001}
{DeForest}, C.~E., {Plunkett}, S.~P., \& {Andrews}, M.~D. 2001{\natexlab{b}},
  \bibinfo{title}{{Observation of Polar Plumes at High Solar Altitudes},} \apj,
  546, 569, \dodoi{10.1086/318221}

% type= article
\bibitem[{G. {Del Zanna} {et~al.}(2003){Del Zanna}, {Bromage}, \&
  {Mason}}]{delzanna-2003}
{Del Zanna}, G., {Bromage}, B.~J.~I., \& {Mason}, H.~E. 2003,
  \bibinfo{title}{{Spectroscopic characteristics of polar plumes},} \aap, 398,
  743, \dodoi{10.1051/0004-6361:20021628}

% type= article
\bibitem[{L. {Del Zanna} {et~al.}(1998){Del Zanna}, {von Steiger}, \&
  {Velli}}]{delzanna-1998}
{Del Zanna}, L., {von Steiger}, R., \& {Velli}, M. 1998, \bibinfo{title}{{The
  Expansion of Coronal Plumes in the Fast Solar Wind},} \ssr, 85, 349,
  \dodoi{10.1023/A:1005127206950}

% type= article
\bibitem[{B.~N. {Dwivedi} \& K. {Wilhelm}(2015){Dwivedi} \&
  {Wilhelm}}]{dwivedi-2015}
{Dwivedi}, B.~N., \& {Wilhelm}, K. 2015, \bibinfo{title}{{Solar Coronal Plumes
  and the Fast Solar Wind},} Journal of Astrophysics and Astronomy, 36, 185,
  \dodoi{10.1007/s12036-015-9326-0}

% type= article
\bibitem[{R. {Fisher} \& M. {Guhathakurta}(1995){Fisher} \&
  {Guhathakurta}}]{fisher-1995}
{Fisher}, R., \& {Guhathakurta}, M. 1995, \bibinfo{title}{{Physical Properties
  of Polar Coronal Rays and Holes as Observed with the SPARTAN 201-01
  Coronagraph},} \apjl, 447, L139, \dodoi{10.1086/309582}

% type= article
\bibitem[{A.~H. {Gabriel} {et~al.}(2003){Gabriel}, {Bely-Dubau}, \&
  {Lemaire}}]{gabriel-2003}
{Gabriel}, A.~H., {Bely-Dubau}, F., \& {Lemaire}, P. 2003, \bibinfo{title}{{The
  Contribution of Polar Plumes to the Fast Solar Wind},} \apj, 589, 623,
  \dodoi{10.1086/374416}

% type= article
\bibitem[{J. {Gorman} {et~al.}(2022){Gorman}, {Chitta}, \&
  {Peter}}]{gorman-2022}
{Gorman}, J., {Chitta}, L.~P., \& {Peter}, H. 2022,
  \bibinfo{title}{{Spectroscopic observation of a transition region network
  jet},} \aap, 660, A116, \dodoi{10.1051/0004-6361/202142995}

% type= article
\bibitem[{J.~W. {Harvey}(1965){Harvey}}]{harvey-1965}
{Harvey}, J.~W. 1965, \bibinfo{title}{{Coronal Polar Rays and Polar Magnetic
  Fields.},} \apj, 141, 832, \dodoi{10.1086/148175}

% type= article
\bibitem[{D.~M. {Hassler} {et~al.}(1997){Hassler}, {Wilhelm}, {Lemaire}, \&
  {Sch{\"u}hle}}]{hassler-1997}
{Hassler}, D.~M., {Wilhelm}, K., {Lemaire}, P., \& {Sch{\"u}hle}, U. 1997,
  \bibinfo{title}{{Observations of Polar Plumes with the SUMER Instrument on
  SOHO},} \solphys, 175, 375, \dodoi{10.1023/A:1004959324214}

% type= article
\bibitem[{R. {Hosseini} {et~al.}(2024){Hosseini}, {Kayshap}, {Alipour}, \&
  {Safari}}]{hosseini-2024}
{Hosseini}, R., {Kayshap}, P., {Alipour}, N., \& {Safari}, H. 2024,
  \bibinfo{title}{{Asymmetry of the spectral lines of the coronal hole and
  quiet Sun in the transition region},} \mnras, 529, 3424,
  \dodoi{10.1093/mnras/stae356}

% type= article
\bibitem[{F. {Jiao} {et~al.}(2015){Jiao}, {Xia}, {Li}, {Huang}, {Li},
  {Chandrashekhar}, {Mou}, \& {Fu}}]{jiao-2015}
{Jiao}, F., {Xia}, L., {Li}, B., {et~al.} 2015, \bibinfo{title}{{Sources of
  Quasi-periodic Propagating Disturbances above a Solar Polar Coronal Hole},}
  \apjl, 809, L17, \dodoi{10.1088/2041-8205/809/1/L17}

% type= article
\bibitem[{P. {Kayshap} {et~al.}(2018){Kayshap}, {Murawski}, {Srivastava}, \&
  {Dwivedi}}]{kayshap-2018}
{Kayshap}, P., {Murawski}, K., {Srivastava}, A.~K., \& {Dwivedi}, B.~N. 2018,
  \bibinfo{title}{{Rotating network jets in the quiet Sun as observed by
  IRIS},} \aap, 616, A99, \dodoi{10.1051/0004-6361/201730990}

% type= article
\bibitem[{M. {Koletti} {et~al.}(2024){Koletti}, {Gontikakis}, {Patsourakos}, \&
  {Tsinganos}}]{koletti-2024}
{Koletti}, M., {Gontikakis}, C., {Patsourakos}, S., \& {Tsinganos}, K. 2024,
  \bibinfo{title}{{Multiwavelength study of on-disk coronal-hole jets with IRIS
  and SDO observations},} \aap, 690, A11, \dodoi{10.1051/0004-6361/202348446}

% type= inproceedings
\bibitem[{P. {Lamy} {et~al.}(1997){Lamy}, {Liebaria}, {Koutchmy}, {Reynet},
  {Molodensky}, {Howard}, {Schwenn}, \& {Simnett}}]{lamy-1997}
{Lamy}, P., {Liebaria}, A., {Koutchmy}, S., {et~al.} 1997,
  \bibinfo{title}{{Characterisation of Polar Plumes from LASCO-C2 Images in
  Early 1996},} in ESA Special Publication, Vol. 404, Fifth SOHO Workshop: The
  Corona and Solar Wind Near Minimum Activity, ed. A.~{Wilson}, 487

% type= article
\bibitem[{J. {Leenaarts} {et~al.}(2013{\natexlab{a}}){Leenaarts}, {Pereira},
  {Carlsson}, {Uitenbroek}, \& {De Pontieu}}]{leenaarts-2013}
{Leenaarts}, J., {Pereira}, T.~M.~D., {Carlsson}, M., {Uitenbroek}, H., \& {De
  Pontieu}, B. 2013{\natexlab{a}}, \bibinfo{title}{{The Formation of IRIS
  Diagnostics. II. The Formation of the Mg II h\&k Lines in the Solar
  Atmosphere},} \apj, 772, 90, \dodoi{10.1088/0004-637X/772/2/90}

% type= article
\bibitem[{J. {Leenaarts} {et~al.}(2013{\natexlab{b}}){Leenaarts}, {Pereira},
  {Carlsson}, {Uitenbroek}, \& {De Pontieu}}]{leenaarts-2013a}
{Leenaarts}, J., {Pereira}, T.~M.~D., {Carlsson}, M., {Uitenbroek}, H., \& {De
  Pontieu}, B. 2013{\natexlab{b}}, \bibinfo{title}{{The Formation of IRIS
  Diagnostics. I. A Quintessential Model Atom of Mg II and General Formation
  Properties of the Mg II h\&k Lines},} \apj, 772, 89,
  \dodoi{10.1088/0004-637X/772/2/89}

% type= article
\bibitem[{J.~R. {Lemen} {et~al.}(2012){Lemen}, {Title}, {Akin}, {Boerner},
  {Chou}, {Drake}, {Duncan}, {Edwards}, {Friedlaender}, {Heyman}, {Hurlburt},
  {Katz}, {Kushner}, {Levay}, {Lindgren}, {Mathur}, {McFeaters}, {Mitchell},
  {Rehse}, {Schrijver}, {Springer}, {Stern}, {Tarbell}, {Wuelser}, {Wolfson},
  {Yanari}, {Bookbinder}, {Cheimets}, {Caldwell}, {Deluca}, {Gates}, {Golub},
  {Park}, {Podgorski}, {Bush}, {Scherrer}, {Gummin}, {Smith}, {Auker},
  {Jerram}, {Pool}, {Soufli}, {Windt}, {Beardsley}, {Clapp}, {Lang}, \&
  {Waltham}}]{aia-lemen}
{Lemen}, J.~R., {Title}, A.~M., {Akin}, D.~J., {et~al.} 2012,
  \bibinfo{title}{{The Atmospheric Imaging Assembly (AIA) on the Solar Dynamics
  Observatory (SDO)},} \solphys, 275, 17, \dodoi{10.1007/s11207-011-9776-8}

% type= article
\bibitem[{B. {Malaker} {et~al.}(2024){Malaker}, {Upendran}, \&
  {Tripathi}}]{malaker-2024}
{Malaker}, B., {Upendran}, V., \& {Tripathi}, D. 2024,
  \bibinfo{title}{{Thermodynamic Evolution of Plumes},} \apj, 974, 163,
  \dodoi{10.3847/1538-4357/ad6c4b}

% type= article
\bibitem[{J. {Mart{\'\i}nez-Sykora} {et~al.}(2018){Mart{\'\i}nez-Sykora}, {De
  Pontieu}, {De Moortel}, {Hansteen}, \& {Carlsson}}]{martnezsykora-2018}
{Mart{\'\i}nez-Sykora}, J., {De Pontieu}, B., {De Moortel}, I., {Hansteen},
  V.~H., \& {Carlsson}, M. 2018, \bibinfo{title}{{Impact of Type II Spicules in
  the Corona: Simulations and Synthetic Observables},} \apj, 860, 116,
  \dodoi{10.3847/1538-4357/aac2ca}

% type= article
\bibitem[{J. {Mart{\'\i}nez-Sykora} {et~al.}(2017){Mart{\'\i}nez-Sykora}, {De
  Pontieu}, {Hansteen}, {Rouppe van der Voort}, {Carlsson}, \&
  {Pereira}}]{martnezsykora-2017}
{Mart{\'\i}nez-Sykora}, J., {De Pontieu}, B., {Hansteen}, V.~H., {et~al.} 2017,
  \bibinfo{title}{{On the generation of solar spicules and Alfv{\'e}nic
  waves},} Science, 356, 1269, \dodoi{10.1126/science.aah5412}

% type= article
\bibitem[{J. {Mart{\'\i}nez-Sykora} {et~al.}(2020){Mart{\'\i}nez-Sykora},
  {Leenaarts}, {De Pontieu}, {N{\'o}brega-Siverio}, {Hansteen}, {Carlsson}, \&
  {Szydlarski}}]{martnezsykora-2020}
{Mart{\'\i}nez-Sykora}, J., {Leenaarts}, J., {De Pontieu}, B., {et~al.} 2020,
  \bibinfo{title}{{Ion-neutral Interactions and Nonequilibrium Ionization in
  the Solar Chromosphere},} \apj, 889, 95, \dodoi{10.3847/1538-4357/ab643f}

% type= article
\bibitem[{R.~J. {Morton} {et~al.}(2023){Morton}, {Sharma}, {Tajfirouze}, \&
  {Miriyala}}]{morton-2023a}
{Morton}, R.~J., {Sharma}, R., {Tajfirouze}, E., \& {Miriyala}, H. 2023,
  \bibinfo{title}{{Alfv{\'e}nic waves in the inhomogeneous solar atmosphere},}
  Reviews of Modern Plasma Physics, 7, 17, \dodoi{10.1007/s41614-023-00118-3}

% type= article
\bibitem[{R.~J. {Morton} {et~al.}(2015){Morton}, {Tomczyk}, \&
  {Pinto}}]{morton-2015}
{Morton}, R.~J., {Tomczyk}, S., \& {Pinto}, R. 2015,
  \bibinfo{title}{{Investigating Alfv{\'e}nic wave propagation in coronal
  open-field regions},} Nature Communications, 6, 7813,
  \dodoi{10.1038/ncomms8813}

% type= article
\bibitem[{N. {Narang} {et~al.}(2016){Narang}, {Arbacher}, {Tian}, {Banerjee},
  {Cranmer}, {DeLuca}, \& {McKillop}}]{narang-2016}
{Narang}, N., {Arbacher}, R.~T., {Tian}, H., {et~al.} 2016,
  \bibinfo{title}{{Statistical Study of Network Jets Observed in the Solar
  Transition Region: a Comparison Between Coronal Holes and Quiet-Sun
  Regions},} \solphys, 291, 1129, \dodoi{10.1007/s11207-016-0886-1}

% type= article
\bibitem[{G. {Newkirk} \& J. {Harvey}(1968){Newkirk} \&
  {Harvey}}]{newkirk-1968}
{Newkirk}, Jr., G., \& {Harvey}, J. 1968, \bibinfo{title}{{Coronal Polar
  Plumes},} \solphys, 3, 321, \dodoi{10.1007/BF00155166}

% type= article
\bibitem[{B. {O'Dwyer} {et~al.}(2010){O'Dwyer}, {Del Zanna}, {Mason}, {Weber},
  \& {Tripathi}}]{o'dwyer-2010}
{O'Dwyer}, B., {Del Zanna}, G., {Mason}, H.~E., {Weber}, M.~A., \& {Tripathi},
  D. 2010, \bibinfo{title}{{SDO/AIA response to coronal hole, quiet Sun, active
  region, and flare plasma},} \aap, 521, A21,
  \dodoi{10.1051/0004-6361/201014872}

% type= article
\bibitem[{N.~K. {Panesar} {et~al.}(2018){Panesar}, {Sterling}, {Moore},
  {Tiwari}, {De Pontieu}, \& {Norton}}]{panesar-2018}
{Panesar}, N.~K., {Sterling}, A.~C., {Moore}, R.~L., {et~al.} 2018,
  \bibinfo{title}{{IRIS and SDO Observations of Solar Jetlets Resulting from
  Network-edge Flux Cancelation},} \apjl, 868, L27,
  \dodoi{10.3847/2041-8213/aaef37}

% type= article
\bibitem[{V. {Pant} {et~al.}(2015){Pant}, {Dolla}, {Mazumder}, {Banerjee},
  {Krishna Prasad}, \& {Panditi}}]{pant-2014}
{Pant}, V., {Dolla}, L., {Mazumder}, R., {et~al.} 2015,
  \bibinfo{title}{{Dynamics of On-disk Plumes as Observed with the Interface
  Region Imaging Spectrograph, the Atmospheric Imaging Assembly, and the
  Helioseismic and Magnetic Imager},} \apj, 807, 71,
  \dodoi{10.1088/0004-637X/807/1/71}

% type= article
\bibitem[{T.~M.~D. {Pereira} {et~al.}(2013){Pereira}, {Leenaarts}, {De
  Pontieu}, {Carlsson}, \& {Uitenbroek}}]{pereira_2013}
{Pereira}, T.~M.~D., {Leenaarts}, J., {De Pontieu}, B., {Carlsson}, M., \&
  {Uitenbroek}, H. 2013, \bibinfo{title}{{The Formation of IRIS Diagnostics.
  III. Near-ultraviolet Spectra and Images},} \apj, 778, 143,
  \dodoi{10.1088/0004-637X/778/2/143}

% type= article
\bibitem[{W.~D. {Pesnell} {et~al.}(2012){Pesnell}, {Thompson}, \&
  {Chamberlin}}]{sdo-pesnell}
{Pesnell}, W.~D., {Thompson}, B.~J., \& {Chamberlin}, P.~C. 2012,
  \bibinfo{title}{{The Solar Dynamics Observatory (SDO)},} \solphys, 275, 3,
  \dodoi{10.1007/s11207-011-9841-3}

% type= article
\bibitem[{S. {Pucci} {et~al.}(2014){Pucci}, {Poletto}, {Sterling}, \&
  {Romoli}}]{pucci-2014}
{Pucci}, S., {Poletto}, G., {Sterling}, A.~C., \& {Romoli}, M. 2014,
  \bibinfo{title}{{Birth, Life, and Death of a Solar Coronal Plume},} \apj,
  793, 86, \dodoi{10.1088/0004-637X/793/2/86}

% type= article
\bibitem[{N.~E. {Raouafi} {et~al.}(2008){Raouafi}, {Petrie}, {Norton},
  {Henney}, \& {Solanki}}]{raouafi-2008}
{Raouafi}, N.~E., {Petrie}, G.~J.~D., {Norton}, A.~A., {Henney}, C.~J., \&
  {Solanki}, S.~K. 2008, \bibinfo{title}{{Evidence for Polar Jets as Precursors
  of Polar Plume Formation},} \apjl, 682, L137, \dodoi{10.1086/591125}

% type= article
\bibitem[{N.~E. {Raouafi} \& G. {Stenborg}(2014){Raouafi} \&
  {Stenborg}}]{raouafi-2014}
{Raouafi}, N.~E., \& {Stenborg}, G. 2014, \bibinfo{title}{{Role of Transients
  in the Sustainability of Solar Coronal Plumes},} \apj, 787, 118,
  \dodoi{10.1088/0004-637X/787/2/118}

% type= article
\bibitem[{K. {Saito}(1958){Saito}}]{saito-1958}
{Saito}, K. 1958, \bibinfo{title}{{Polar Rays of the Solar Corona},} \pasj, 10,
  49, \dodoi{10.1093/pasj/10.2.49}

% type= article
\bibitem[{K. {Saito}(1965){Saito}}]{saito-1965}
{Saito}, K. 1965, \bibinfo{title}{{Polar Rays of the Solar Corona, II.},}
  \pasj, 17, 1

% type= article
\bibitem[{T. {Samanta} {et~al.}(2015){Samanta}, {Pant}, \&
  {Banerjee}}]{samanta-2015}
{Samanta}, T., {Pant}, V., \& {Banerjee}, D. 2015, \bibinfo{title}{{Propagating
  Disturbances in the Solar Corona and Spicular Connection},} \apjl, 815, L16,
  \dodoi{10.1088/2041-8205/815/1/L16}

% type= article
\bibitem[{J. {Schou} {et~al.}(2012){Schou}, {Scherrer}, {Bush}, {Wachter},
  {Couvidat}, {Rabello-Soares}, {Bogart}, {Hoeksema}, {Liu}, {Duvall}, {Akin},
  {Allard}, {Miles}, {Rairden}, {Shine}, {Tarbell}, {Title}, {Wolfson},
  {Elmore}, {Norton}, \& {Tomczyk}}]{hmi-schou}
{Schou}, J., {Scherrer}, P.~H., {Bush}, R.~I., {et~al.} 2012,
  \bibinfo{title}{{Design and Ground Calibration of the Helioseismic and
  Magnetic Imager (HMI) Instrument on the Solar Dynamics Observatory (SDO)},}
  \solphys, 275, 229, \dodoi{10.1007/s11207-011-9842-2}

% type= inproceedings
\bibitem[{T. {Shimizu} {et~al.}(2019){Shimizu}, {Imada}, {Kawate}, {Ichimoto},
  {Suematsu}, {Hara}, {Katsukawa}, {Kubo}, {Toriumi}, {Watanabe}, {Yokoyama},
  {Korendyke}, {Warren}, {Tarbell}, {De Pontieu}, {Teriaca}, {Sch{\"u}hle},
  {Solanki}, {Harra}, {Matthews}, {Fludra}, {Auch{\`e}re}, {Andretta},
  {Naletto}, \& {Zhukov}}]{solarc}
{Shimizu}, T., {Imada}, S., {Kawate}, T., {et~al.} 2019, \bibinfo{title}{{The
  Solar-C\_EUVST mission},} in Society of Photo-Optical Instrumentation
  Engineers (SPIE) Conference Series, Vol. 11118, UV, X-Ray, and Gamma-Ray
  Space Instrumentation for Astronomy XXI, ed. O.~H. {Siegmund}, 1111807,
  \dodoi{10.1117/12.2528240}

% type= article
\bibitem[{H. {Tian} {et~al.}(2014){Tian}, {DeLuca}, {Cranmer}, {De Pontieu},
  {Peter}, {Mart{\'\i}nez-Sykora}, {Golub}, {McKillop}, {Reeves}, {Miralles},
  {McCauley}, {Saar}, {Testa}, {Weber}, {Murphy}, {Lemen}, {Title}, {Boerner},
  {Hurlburt}, {Tarbell}, {Wuelser}, {Kleint}, {Kankelborg}, {Jaeggli},
  {Carlsson}, {Hansteen}, \& {McIntosh}}]{tian-2014a}
{Tian}, H., {DeLuca}, E.~E., {Cranmer}, S.~R., {et~al.} 2014,
  \bibinfo{title}{{Prevalence of small-scale jets from the networks of the
  solar transition region and chromosphere},} Science, 346, 1255711,
  \dodoi{10.1126/science.1255711}

% type= article
\bibitem[{V. {Upendran} \& D. {Tripathi}(2021){Upendran} \&
  {Tripathi}}]{Upendran_2021_C2}
{Upendran}, V., \& {Tripathi}, D. 2021, \bibinfo{title}{{Properties of the C II
  1334 {\r{A}} Line in Coronal Hole and Quiet Sun as Observed by IRIS},} \apj,
  922, 112, \dodoi{10.3847/1538-4357/ac2575}

% type= article
\bibitem[{V. {Upendran} \& D. {Tripathi}(2022){Upendran} \&
  {Tripathi}}]{upendran_2022_Mg2}
{Upendran}, V., \& {Tripathi}, D. 2022, \bibinfo{title}{{On the Formation of
  Solar Wind and Switchbacks, and Quiet Sun Heating},} \apj, 926, 138,
  \dodoi{10.3847/1538-4357/ac3d88}

% type= article
\bibitem[{V. {Upendran} {et~al.}(2025){Upendran}, {Tripathi}, {Vaidya},
  {Cheung}, \& {Yokoyama}}]{Upendran_2025_sim}
{Upendran}, V., {Tripathi}, D., {Vaidya}, B., {Cheung}, M. C.~M., \&
  {Yokoyama}, T. 2025, \bibinfo{title}{{Comparison of Plasma Dynamics in
  Coronal Holes and Quiet Sun Using Flux Emergence Simulations},} \apj, 985,
  27, \dodoi{10.3847/1538-4357/adc5fd}

% type= article
\bibitem[{V.~M. {Uritsky} {et~al.}(2021){Uritsky}, {DeForest}, {Karpen},
  {DeVore}, {Kumar}, {Raouafi}, \& {Wyper}}]{uritsky-2021}
{Uritsky}, V.~M., {DeForest}, C.~E., {Karpen}, J.~T., {et~al.} 2021,
  \bibinfo{title}{{Plumelets: Dynamic Filamentary Structures in Solar Coronal
  Plumes},} \apj, 907, 1, \dodoi{10.3847/1538-4357/abd186}

% type= article
\bibitem[{H.~C. {van de Hulst}(1950){van de Hulst}}]{vandehulst-1950}
{van de Hulst}, H.~C. 1950, \bibinfo{title}{{On the polar rays of the corona
  (Errata: 11 VIII)},} \bain, 11, 150

% type= article
\bibitem[{M. {Velli} {et~al.}(2011){Velli}, {Lionello}, {Linker}, \&
  {Miki{\'c}}}]{velli-2011}
{Velli}, M., {Lionello}, R., {Linker}, J.~A., \& {Miki{\'c}}, Z. 2011,
  \bibinfo{title}{{Coronal Plumes in the Fast Solar Wind},} \apj, 736, 32,
  \dodoi{10.1088/0004-637X/736/1/32}

% type= article
\bibitem[{P. Virtanen {et~al.}(2020)Virtanen, Gommers, Oliphant, Haberland,
  Reddy, Cournapeau, Burovski, Peterson, Weckesser, Bright, {van der Walt},
  Brett, Wilson, Millman, Mayorov, Nelson, Jones, Kern, Larson, Carey, Polat,
  Feng, Moore, {VanderPlas}, Laxalde, Perktold, Cimrman, Henriksen, Quintero,
  Harris, Archibald, Ribeiro, Pedregosa, {van Mulbregt}, \& {SciPy 1.0
  Contributors}}]{scipy}
Virtanen, P., Gommers, R., Oliphant, T.~E., {et~al.} 2020,
  \bibinfo{title}{{{SciPy} 1.0: Fundamental Algorithms for Scientific Computing
  in Python},} Nature Methods, 17, 261, \dodoi{10.1038/s41592-019-0686-2}

% type= article
\bibitem[{A.~B.~C. {Walker} {et~al.}(1993){Walker}, {Deforest}, {Hoover}, \&
  {Barbee}}]{walker-1993}
{Walker}, Jr., A.~B.~C., {Deforest}, C.~E., {Hoover}, R.~B., \& {Barbee}, Jr.,
  T.~W. 1993, \bibinfo{title}{{Thermal and Density Structure of Polar Plumes},}
  \solphys, 148, 239, \dodoi{10.1007/BF00645089}

% type= article
\bibitem[{Y.~M. {Wang}(1994){Wang}}]{wang-1994}
{Wang}, Y.~M. 1994, \bibinfo{title}{{Polar Plumes and the Solar Wind},} \apjl,
  435, L153, \dodoi{10.1086/187617}

% type= article
\bibitem[{Y.~M. {Wang} \& N.~R. {Sheeley}(1995){Wang} \& {Sheeley}}]{wang-1995}
{Wang}, Y.~M., \& {Sheeley}, Jr., N.~R. 1995, \bibinfo{title}{{Coronal Plumes
  and Their Relationship to Network Activity},} \apj, 452, 457,
  \dodoi{10.1086/176317}

% type= article
\bibitem[{Y.~M. {Wang} {et~al.}(2016){Wang}, {Warren}, \&
  {Muglach}}]{wang-2016}
{Wang}, Y.~M., {Warren}, H.~P., \& {Muglach}, K. 2016,
  \bibinfo{title}{{Converging Supergranular Flows and the Formation of Coronal
  Plumes},} \apj, 818, 203, \dodoi{10.3847/0004-637X/818/2/203}

% type= article
\bibitem[{A. {Weitz} {et~al.}(2025){Weitz}, {Tiwari}, {Cauzzi}, {Reardon}, \&
  {De Pontieu}}]{weitz-2025}
{Weitz}, A., {Tiwari}, S.~K., {Cauzzi}, G., {Reardon}, K.~P., \& {De Pontieu},
  B. 2025, \bibinfo{title}{{Bright Dots and Coronal Plume Formation in Sunspot
  Penumbra},} \apj, 988, 133, \dodoi{10.3847/1538-4357/ade3c1}

% type= article
\bibitem[{K. {Wilhelm} \& R. {Bodmer}(1998){Wilhelm} \&
  {Bodmer}}]{wilhelm-1998a}
{Wilhelm}, K., \& {Bodmer}, R. 1998, \bibinfo{title}{{Solar EUV and UV Emission
  Line Observations Above a Polar Coronal Hole},} \ssr, 85, 371,
  \dodoi{10.1023/A:1005187509676}

% type= article
\bibitem[{K. {Wilhelm} {et~al.}(1998){Wilhelm}, {Marsch}, {Dwivedi}, {Hassler},
  {Lemaire}, {Gabriel}, \& {Huber}}]{wilhelm-1998}
{Wilhelm}, K., {Marsch}, E., {Dwivedi}, B.~N., {et~al.} 1998,
  \bibinfo{title}{{The Solar Corona Above Polar Coronal Holes as Seen by SUMER
  on SOHO},} \apj, 500, 1023, \dodoi{10.1086/305756}

% type= incollection
\bibitem[{G.~L. {Withbroe} {et~al.}(1991){Withbroe}, {Feldman}, \&
  {Ahluwalia}}]{withbroe-1991}
{Withbroe}, G.~L., {Feldman}, W.~C., \& {Ahluwalia}, H.~S. 1991,
  \bibinfo{title}{{The solar wind and its coronal origins.},} in Solar Interior
  and Atmosphere, ed. A.~N. {Cox}, W.~C. {Livingston}, \& M.~S. {Matthews},
  1087--1106

\end{thebibliography}
\end{document}